\documentstyle[11pt,aaspp4]{article}
\def\ges{\;_\sim^>\;}
\def\les{\;_\sim^<\;}
 
\lefthead{Guerra \& Daly}
\righthead{Outflows of AGN}
 
\begin{document}
 
\title{Outflow Angles, Bulk Lorentz Factors, and Kinematics of 
Outflows from the Cores of AGN}

\author{Erick J. Guerra and Ruth A. Daly\altaffilmark{1}}
\affil{Department of Physics, Joseph Henry Laboratories,\\ Princeton
University, Princeton, NJ 08544}
 
\altaffiltext{1}{National Young Investigator}
 
\begin{abstract}

Outflow angles and bulk Lorentz factors for 43 sources that 
have proper motions compiled by Vermeulen \& Cohen (1994)
are computed on the basis of Doppler factors and
observed apparent motions in the plane of the sky.  These estimates of
outflow angles and bulk Lorentz factors are discussed 
along with their agreement
with orientation unified models of AGN.

Intrinsic (i.e. rest frame)
brightness temperatures computed by using the
inverse Compton and equipartition Doppler factors
are discussed along with their relevance to the ``Inverse Compton 
catastrophe''.
Intrinsic luminosity densities and luminosities are presented, and the
role of systematic errors is discussed. 

These studies are carried out using a sample of 100 compact 
radio sources compiled by Ghisellini et al. (1993).  Error estimates for
previously computed inverse Compton Doppler 
factors and equipartition Doppler factors are presented for these sources,
along with a few updates of these Doppler factor estimates. 

\end{abstract}
 

\section{INTRODUCTION}

It is generally accepted that relativistic outflows from the core regions of 
radio-loud active galactic
nuclei (AGN) are responsible for many of the 
interesting phenomena that are observed in these sources.
Relativistic motion of synchrotron emitting plasma will result in
the Doppler boosting of the synchrotron radiation from these outflows 
(discussed by many authors, e.g. Marscher 1987).
In addition, relativistic outflows would 
readily explain the apparent superluminal motion observed on VLBI (Very Long 
Baseline Interferometry) scales (e.g. Pearson \& Zensus 1987).

The Doppler factor is a key quantity in deriving intrinsic
properties of an AGN.  It is well known that the unbeamed
flux density of a radiating source is related to
the observed flux density 
by the Doppler factor.
When
combined with the apparent motion in the plane of the sky,
the Doppler factor can be used to estimate the Lorentz factor
and the viewing angle of an outflow, 
assuming the motion responsible for the Doppler boosting
has the same speed as the motion of radio features in the plane of the sky 
(Ghisellini et al. 1993; Daly, Guerra, \& G\"uijosa 1997).  The 
values of these Lorentz factors
and viewing angles are
significant in
understanding AGN physics and are relevant to 
orientation unified
models of AGN.

There are several methods for computing the
Doppler factor.  One method, called
the inverse Compton Doppler factor, is
derived by assuming the observed X-ray flux is caused by inverse Compton
scattering of synchrotron photons off the radiating particles (e.g.
Marscher 1987).  A second method, called the equipartition
Doppler factor, is derived assuming that the
sources are at or near equipartition of energy between radiating
particles and magnetic field  (Readhead 1994).  Other 
methods (not discussed further here)
include those based on variability
measurements (Rees 1967) or jet/counter jet flux ratios (Conway 1982).
Ghisellini et al. (1993; hereafter GPCM93) compute
the inverse Compton Doppler factor for a sample of 105 sources, using
data compiled from the literature.  The data
compiled by GPCM93 were used by G\"uijosa \& Daly (1996) to compute 
equipartition Doppler factors.
These sets of inverse Compton Doppler factors and equipartition
Doppler factors are used here to compute
the Lorentz factors, viewing angles, and
intrinsic properties of outflows from these sources.

Updated estimates of both inverse Compton and equipartition Doppler factors 
for a sample of 100 sources
are discussed in \S \ref{ssec:GPCM}; intrinsic brightness 
temperatures computed for both sets of Doppler factors are discussed
in \S 
\ref{ssec:temp}.
The reliability of Doppler factors
and the errors assigned to these estimates
are discussed in \S \ref{ssec:rely} and
\S \ref{ssec:err} 
respectively.  Estimates of the intrinsic
luminosity density and intrinsic luminosity are examined in \S
\ref{ssec:lum}.  The computations
of the bulk Lorentz factor and
the viewing angle 
are discussed in \S \ref{sec:anres}.  A subsample of sources with estimates
of Doppler factors and apparent motion in the plane of the
sky is introduced in \S \ref{ssec:VC}, and a set of solutions for
the bulk Lorentz factor and viewing angle are computed
in \S \ref{ssec:solve} for this subsample.
In \S \ref{ssec:comm}, these results are discussed
in the context of orientation
unified models.  A general discussion follows in \S
\ref{sec:disc}.

\section{A SAMPLE OF DOPPLER FACTOR ESTIMATES \label{sec:est}}

\subsection{Previous and Updated Doppler Factor Estimates \label{ssec:GPCM}}

GPCM93 assemble the relevant data to compute the inverse Compton
Doppler factor for 
those AGN with VLBI core size data available in the literature circa
1986-92.  The five BL Lacs without known redshifts are excluded from
the sample examined here, since inverse Compton and equipartition
Doppler factor estimates depend on
redshift (see eqs. \ref{eq:dic} \& \ref{eq:deq}).  This leaves 100 sources
in this sample which fall into the following classifications (GPCM93): 
32 BL Lacerate objects (BL Lac),
53 core-dominated quasars (CDQ), 11
lobe-dominated quasars (LDQ), and 9 radio galaxies (RG).  The CDQs are
further classified on the basis of their optical polarization
into 24 core-dominated high-polarization quasars (CDHPQ),
22 core-dominated low-polarization quasars (CDLPQ), and 7
core-dominated
quasars without polarization information (CDQ-NPI).
A detailed discussion of these classifications can be found in GPCM93.

Both the inverse Compton Doppler factor, $\delta_{IC}$, and the equipartition 
Doppler factor, 
$\delta_{eq}$, can be computed using
the data compiled by GPCM93 (see the Appendix for the relevant formulae).
By assuming that the observed radio frequencies and radio
flux densities are the
true peak radio frequencies and radio flux densities, GPCM93 use the data 
to compute $\delta_{IC}$.  This approximation is made
since the multi-frequency observations and analysis of the radio spectra
necessary to obtain true peak frequencies and flux densities are available
for only a small fraction of these
sources (e.g. Marscher \& Broderick 1985, Unwin et al. 1994).
G\"uijosa \& Daly (1996) recompute $\delta_{IC}$ taking into account
corrections to the angular size and flux density neglected by GPCM93 (see
Appendix),
and compute $\delta_{eq}$ with the same assumptions
for this sample.

Both $\delta_{eq}$ and $\delta_{IC}$ are recomputed in this study assuming
a deceleration parameter of $q_o=0.05$ (i.e. $\Omega_o=0.1$,
$\Omega_{\Lambda}=0$) and $h=0.75$, and using 
updated VLBI data for four sources (1101+384, 0615+820, 1039+811, \& 
1150+812)
from Xu et al. (1995); here $\Omega_o$ is
the mean mass density relative to the critical value, $\Omega_{\Lambda}$ is 
the normalized cosmological constant, and Hubble's constant 
is parameterized in the usual way: 
$H_o = 100\;h\mbox{ km s}^{-1}\mbox{ kpc}^{-1}$.  
When computing $\delta_{eq}$ and $\delta_{IC}$ for each source,
we assume an optically thin spectral index $\alpha = 0.75$ ($S_{\nu}
\propto\nu^{-\alpha}$) and a spherical geometry, as was done by GPCM93 
and G\"uijosa \& Daly (1996) (see eqs. \ref{eq:dic} \& \ref{eq:deq}).
Note that 
$\delta_{eq}$ is changed by
only a few percent relative to the values computed assuming an Einstein
de-Sitter universe ($\Omega_o=1.0$, $\Omega_{\Lambda}=0$) 
and $h=1$, while $\delta_{IC}$
is independent of these assumptions.  
Following GPCM93,
upper bounds on angular sizes and lower bounds on redshifts are taken as
detections in this section,
but are treated as bounds when computing 
bulk Lorentz factors and viewing angles
in \S \ref{ssec:solve}.

Figures \ref{fig:deleq}a \& \ref{fig:deleq}b show $\delta_{eq}$ and
$\delta_{IC}$ respectively, computed as described in the Appendix,
as functions of $(1+z)$.
In these figures and throughout, solid circles represent BL Lacs,
solid diamonds represent CDHPQs, solid squares represent CDLPQs,
solid triangles represent CDQ-NPIs, open diamonds represent LDQs,
and open squares represent RGs.
The error estimates for $\delta_{eq}$ and $\delta_{IC}$
are discussed in \S \ref{ssec:err}.
The trend in redshift is easily understood in terms of the effect of
Doppler boosting on the observed flux density and the flux limited nature
of this sample.  Only at low redshift are we able to observe those sources
with lower Doppler factors, and 
the sample is dominated by higher Doppler
factors at all redshifts.  It is also apparent 
that most LDQs and RGs have Doppler factors
less than 1, and most CDQs have Doppler factors greater than one.  BL Lacs have
a much wider range of Doppler factors, covering the entire range of the whole
sample.  

Figure \ref{fig:dophist} shows the distribution of 
the equipartition Doppler factors, 
$\delta_{eq}$ (dotted line), and the inverse Compton Doppler factors,
$\delta_{IC}$ (dashed line), for all 100
sources in the sample; the horizontal axis 
is expressed in terms of $\log\delta$.  Note that the
distributions of $\delta_{eq}$ and $\delta_{IC}$ are similar, and
both have a range of a few orders of magnitude.  In addition,
the peaks of both
distributions are between 5 and 10.
Figures \ref{fig:dophistcl}a-f show the distribution
of $\delta_{eq}$ (dotted line) and $\delta_{IC}$ (dashed line) for each
class of AGN described above.
For BL Lacs, the distributions
range over a few orders of magnitude in $\delta$, with a peak around
$\delta\simeq 5$ (Figure \ref{fig:dophistcl}a).  CDHPQs have 
$\delta_{eq}$ and $\delta_{IC}$ that cover a smaller range of values than BL
Lacs, and peak noticeably at about $\delta \simeq 10$ (Figure
\ref{fig:dophistcl}b), while the
distribution of $\delta_{eq}$ and $\delta_{IC}$ for CDLPQs have a similar
range as CDHPQs, and peak somewhere around $\delta \simeq 5$ (Figure
\ref{fig:dophistcl}c). 
CDQ-NPIs have the same range of
$\delta_{eq}$ and $\delta_{IC}$ as
CDHPQs and CDLPQs
(Figure \ref{fig:dophistcl}d).  The distributions of $\delta_{eq}$ and
$\delta_{IC}$ for LDQs are wider than those for RGs, but both classes
peak at $\delta<1$ (see Figures \ref{fig:dophistcl}e \& \ref{fig:dophistcl}f).

The classification 3C216 as a LDQ should be examined closely since
it has relatively large values for $\delta_{eq}$ and $\delta_{IC}$ 
(61 and 33 respectively).  
The separation of quasars
into CDQ and LDQ was done by GPCM93 on 
the basis of the core dominance parameter
computed using fluxes that were K-corrected to 5 Ghz.  Recent
observation by Reid et al. (1995) at 5 GHz give a core dominance
parameter slightly greater than 1.0, and this would place 3C216 
between the LDQs and CDQs in terms of the core dominance parameter. 
Most LDQs
in this sample exhibit a triple morphology, consisting of a core with a
flat radio spectrum and two lobes with steep radio spectra and have
angular extents of about 10"-30"
(e.g. \cite{hoo}).  3C216 shows a triple morphology, but has an angular
extent of only about 1.7".  It
could be argued that 3C216 should be reassigned as a CDQ, but in this
paper we keep this source in the LDQ sample but make a special note of it.

\subsubsection{Brightness Temperatures \label{ssec:temp}}
Both the equipartition Doppler factor and inverse Compton Doppler factor,
$\delta_{eq}$ and $\delta_{IC}$, are used here to compute
the intrinsic peak
brightness temperature for each source in the sample (see the
Appendix for relevant
formulae).  
Histograms of the observed brightness temperature, intrinsic brightness
temperature base on $\delta_{eq}$, and intrinsic brightness
temperature base on $\delta_{IC}$ ($T_{Bo}$, $T_{Bi}(eq)$,
and $T_{Bi}(IC)$)
are shown in Figure \ref{fig:temphist} for all 100 sources, and in
Figures \ref{fig:temphistcl}a-f for each class of AGN.
Table \ref{tb:bri} lists the mean and the standard deviation of the
mean of
$T_{Bo}$, $T_{Bi}(eq)$, and $T_{Bi}(IC)$ for the full sample and each
class.

The observed peak brightness
temperatures span a few orders of magnitude even within some individual
classes of AGN.  
In contrast, there is a sharp peak in the distribution
of the estimates of
intrinsic brightness temperature, $T_{Bi}(eq)$ and $T_{Bi}(IC)$,
and both are centered around about
$7.5\times10^{10}\hbox{ K}$.  These two 
estimates of the intrinsic brightness temperature agree well with each
other, which
is not surprising given the strong correlation between
$\delta_{eq}$ and $\delta_{IC}$ described above and
by G\"uijosa \& Daly (1996).

It is interesting to note that the mean $T_{Bi}(IC)$
for all classes (except CDQ-NPIs) are consistent within $1\sigma$,
while the LDQs and RGs have larger $T_{Bi}(eq)$ 
than the other
classes at about the $2\sigma$ level.  These two classes have values of
$R=\delta_{eq}/\delta_{IC}$ less than one (G\"uijosa \& Daly 1996), 
which could possibly be due
to a systematic underestimate of $\delta_{eq}$ by about 75\%.
Such an underestimate of $\delta_{eq}$ could account for the larger
$T_{Bi}(eq)$ found for LDQs and RGs.  

The standard
deviations of $T_{Bi}(eq)$ and $T_{Bi}(IC)$ are $0.27\times 10^{11}$ K and
$0.33\times 10^{11}$ K respectively.
It should be noted that $T_{Bi}(eq)\propto S_{op}^{0.1} \nu_{op}^{0.3} 
\theta_d^{0.3}$ and $T_{Bi}(IC)\propto S_{op}^{0} \nu_{op}^{-0.7}
\theta_d^{-0.4}$ which are weak dependences on observable,
much weaker than the dependences of $T_{Bo}$ on the same parameters.
If the range of values
for the observables that go into $T_{Bi}(eq)$ and $T_{Bi}(IC)$
are not too large, then their
narrow distributions can be explained by weak dependences on observable
parameters.  The range of $\theta_d$ contributes the
most to the range of $T_{Bi}(eq)$ and $T_{Bi}(IC)$, since $\nu_{op}$ is 
fixed at discrete values corresponding to the frequencies
of observations, and $S_{op}$ is virtually insignificant
in computing $T_{Bi}(eq)$ and $T_{Bi}(IC)$.
If the narrow ranges of $T_{Bi}(eq)$ 
and $T_{Bi}(IC)$ are due to some physical cause and not the range in 
$\theta_d$, then randomly reassigning the angular sizes used to compute
$T_{Bi}(eq)$ and $T_{Bi}(IC)$ should widen the distributions of each.  
When the angles are reassigned randomly, the means are quite similar
($0.85\times10^{11}$ K and $0.79\times10^{11}$ K respectively),
and the standard
deviations are $0.26\times 10^{11}$ K and
$0.43\times 10^{11}$ K respectively.
Each width is not significantly affected by 
randomly reassigning angular sizes, which suggests that the narrow range
of $T_{Bi}(eq)$ and $T_{Bi}(IC)$ obtained here
is primarily due to the range of observables, and are probably not
due to the physical state of the sources.

Readhead (1994) points out that if the intrinsic brightness temperatures
were cutoff due to the inverse Compton catastrophe
(i.e. the radiation 
lifetime plays a role in the observed cutoff),
one would expect the intrinsic 
brightness temperatures of compact radio sources to
cluster near $10^{11.5}\hbox{ K}$.
Both estimates of intrinsic brightness temperatures for this sample
peak below $10^{11.5}\hbox{ K}$ which agrees with the results of Readhead
(1994).  This suggests that the inverse Compton
catastrophe does not play a role in the observed brightness temperature
distribution, or if the peaks are set by the inverse Compton catastrophe,
a factor of about 4 has not been accounted 
for in computing the unbeamed brightness temperatures.  

\subsection{Reliability of $\delta_{eq}$ and $\delta_{IC}$ 
as Estimators of
the Doppler Factor \label{ssec:rely}}

Previous results indicate that $\delta_{eq}\simeq\delta_{IC}$ for this sample 
(G\"uijosa \& Daly 1996).  To further test this conclusion, the effect of 
randomly reassigning one of the observable quantities used to compute 
$\delta_{eq}$ and $\delta_{IC}$ is examined.  Since both Doppler factors depend
strongly on angular size ($\delta_{eq}\propto\theta_d^{-2.3}$ and $\delta_{IC}
\propto\theta_d^{-1.6}$),
the angular sizes used to compute $\delta_{eq}$
and $\delta_{IC}$ were randomly reassigned between sources within each
class, and the mean values are shown in Table \ref{tb:test}.  
Note that the means
of $\delta_{eq}$ and $\delta_{IC}$ tend to
systematically increase from their true 
values for all classes when the angular sizes are reassigned.

The fractional uncertainties in the mean values 
of $\delta_{eq}$ and $\delta_{IC}$ are of order
20\% to 30\%, so that the mean $\delta_{eq}$ and
$\delta_{IC}$ are significant at about the $3\sigma$ to $5\sigma$ level
(G\"uijosa \& Daly 1996).
In contrast, the ``random'' mean values are significant at $2 \sigma$ or less 
(i.e. their uncertainties are
greater than 50\%, Table \ref{tb:test}).
This suggests that these Doppler factor estimate are reliable.

\subsection{Errors for the Doppler Factor Estimates \label{ssec:err}}

The main source of error for both Doppler factor estimates is the
assumption that the radio flux density and frequency used to compute
the Doppler factor are the peak
radio flux density and peak frequency.  
Thus, an estimate of errors for this data set should focus on
systematic errors instead of statistical ones. 
A dependence of observed angular sizes
on frequency occurs if the
synchrotron-emitting plasma has gradients in magnetic field and particle
density (Marscher 1977, 1987), but since the model
considered here is that of a uniform
sphere, the errors associated with this frequency dependence of
angular size are
neglected.

The computed errors are found by assuming all sources have the same
fractional errors on the peak frequency, $\nu_{op}$;
this uncertainty is likely of order
100\% and dominates all other sources of error, as discussed in 
G\"uijosa \& Daly (1996).  It is assumed that the
flux density has an error which is related to the spectral index
between the observed frequency and the peak frequency, $\alpha_{op}$,
where $S_{op}=S_{o}(\nu_{op}/\nu_{o})^{-\alpha_{op}}$ and
$S_{o}$ and $\nu_{o}$ are the observed flux density and frequency;
$\alpha_{op}$
should not to be confused with the optically thin spectral index, $\alpha$.
In terms of VLBI observed properties,
$\delta_{eq}\propto S_{op}^{1.1}\nu_{op}^{-2.3}\theta_d^{-2.3}$ and
$\delta_{IC}\propto S_{op}^{1.0}\nu_{op}^{-1.3}\theta_d^{-1.6}$, 
assuming $\alpha=0.75$.
These two Doppler factor estimates can be related to
the ratio of the peak frequency to the observed frequency, 
$\delta_{eq}\propto (\nu_{op}/\nu_{o})^{-1.1\alpha_{op}-2.3}$
and $\delta_{IC}\propto
(\nu_{op}/\nu_{o})^{-1.0\alpha_{op}-1.3}$.
One finds that 
\begin{equation}
{{\sigma\delta_{eq}}\over{\delta_{eq}}}=
(1.1\alpha_{op}+2.3)\left({{\sigma\nu_{op}}\over{\nu_{op}}}
\right)\qquad \mbox{and} \qquad
{{\sigma\delta_{IC}}\over{\delta_{IC}}}=
(1.0\alpha_{op}+1.3)\left({{\sigma\nu_{op}}\over{\nu_{op}}}
\right),
\label{eq:deqerr}
\end{equation}
where $\sigma\delta_{eq}$, $\sigma\delta_{IC}$, and $\sigma\nu_{op}$ are
the error estimates for $\delta_{eq}$, $\delta_{IC}$, 
and $\nu_{op}$.

An estimate of $\sigma\nu_{op}/\nu_{op}$ is needed to assign errors for
the Doppler factor estimates using equation (\ref{eq:deqerr}).
A reasonable value of
$\sigma\nu_{op}/\nu_{op}$ is that which gives a reduced $\chi^2$ of
1.0 when fitting $R\equiv\delta_{eq}/\delta_{IC}$ to unity.

Since $R\propto S_{op}^{0.1}\nu_{op}^{-1.0}\theta_d^{-0.7}$ for $\alpha=0.75$,
it can be shown that
$R\propto 
(\nu_{op}/\nu_{o})^{-0.1\alpha_{op}-1.0}$. 
The error on the ratio $R$,
denoted $\sigma R$, can be related to
$\sigma\nu_{op}/\nu_{op}$ by
\begin{equation}
{{\sigma R}\over{R}}=(0.1\alpha_{op}+1.0)
\left({{\sigma\nu_{op}}\over{\nu_{op}}}\right).
\label{eq:Rerr}
\end{equation}
Note that if $\sigma\nu_{op}/\nu_{op}$ and $\alpha_{op}$ are the same
for all sources, then $\sigma R/R$ is the same for all
sources.

The value of $\sigma R/R=0.61$ 
gives a reduced $\chi^2$ of 1.0 when fitting $R$ to a constant equal to
unity for all sources.
A value of $\alpha_{op}=0.4$ is assumed for all 
sources since radio sources with
observed optically
thin spectral indices for their cores have spectral indices in the range
of about 0.5 to 1, and the spectral index must go to zero towards the 
peak.  A lower value of $\alpha_{op}$ could be used, and
this will decrease the errors slightly on $\delta_{eq}$ and
$\delta_{IC}$.  For $\alpha_{op}=0.4$ and $\sigma R/R=0.61$, equation
(\ref{eq:Rerr}) gives $\sigma\nu_{op}/\nu_{op}=0.59$, and one finds that
$\sigma\delta_{eq}/\delta_{eq}=1.6$ and
$\sigma\delta_{IC}/\delta_{IC}=1.0$ using equation (\ref{eq:deqerr}).
These error estimates for $\delta_{eq}$ and $\delta_{IC}$
are adopted throughout this paper.

Some objects in this sample have been observed at multiple frequencies, and 
spectra are available in the literature.  The observed frequencies
used to compute the Doppler factors for eight of the sources
studied here are compared to the peak frequencies
of the integrated spectra available in Xu et al. (1995), in order to check
if our estimate of $\sigma\nu_{op}/\nu_{op}$ is reasonable.  It should
be noted that the integrated spectra include components besides the 
cores, and that the core component is used to
determine the Doppler factor (see \S \ref{sec:disc}).  Here 
we approximate that the
peak of the integrated spectra corresponds to the peak of the core.
On this basis,
four sources have $\sigma\nu_{op}/\nu_{op}$ between 0 and 0.5, and the other
four have $\sigma\nu_{op}/\nu_{op}$ between 0.5 and 4, which would indicate
that the value of $\sigma\nu_{op}/\nu_{op}=0.59$ adopted for the whole sample
is quite reasonable.

\subsection{Luminosity Densities and Luminosities\label{ssec:lum}}

The uncorrected luminosity density is given by 
\begin{equation}
L_{\nu_{un}}=4 \pi (a_or)^2 (1+z) S_o,
\label{eq:oblum}
\end{equation}
where $S_o$ is the observed flux density, $z$ is the source redshift,
$(a_or)$ is the coordinate distance to the source, and $\nu_{un}$ is 
frequency as would be measured in the local Hubble frame of the source without
correcting for Doppler boosting ($\nu_{un}=\nu_o (1+z)$ where $\nu_o$ is the
observed frequency).
The flux density corrected for Doppler boosting of
a uniform spherical source, $S_i$, is given by $S_i=S_o/\delta^3$, where
$\delta$ is the Doppler factor.
The corrected flux density is the flux
density that would be received by an observer
moving with the same velocity with
respect to the Hubble flow as the radiating source.
Substituting $S_i$ for $S_o$ in equation (\ref{eq:oblum}) gives the
Doppler corrected, or intrinsic, luminosity density:
\begin{equation}
L_{\nu i}=4 \pi (a_or)^2 {{1+z}\over{\delta^3}} S_o
\label{eq:inlum}
\end{equation}
where $\nu_i=\nu_o (1+z) / \delta$.
Substituting both estimates of the Doppler factor $\delta_{eq}$ and
$\delta_{IC}$ into equation (\ref{eq:inlum}) gives two sets of intrinsic
luminosity densities which shall be referred to as $L_{\nu i}(eq)$ and
$L_{\nu i}(IC)$ respectively. 

It should be noted that equations (\ref{eq:oblum}) and (\ref{eq:inlum})
are the luminosity densities at frequencies that differ from the
observing frequencies, which are 
$\nu_{un}=\nu_o(1+z)$ and $\nu_i=\nu_o(1+z)/\delta$
respectively.
Thus, the uncorrected luminosity can
be approximated as $L_{un}\approx L_{\nu_{un}}
\nu_o(1+z)$, and the intrinsic luminosity can be approximated as
\begin{equation}
L_{i}\approx L_{\nu i}\nu_o(1+z)/\delta.
\label{eq:inlumint}
\end{equation}
Note
that luminosities computed in this manner
are crude approximations that may be
systematically biased by the approximated peak frequencies, yet still 
give order of magnitude estimates.

Figures \ref{fig:oblumz} \& \ref{fig:oblumintz} show $L_{\nu_{un}}$ and
$L_{un}$ as functions of $(1+z)$.  The flux-limited nature
of this sample is evident in these two figures, and it is quite obvious
that selection effects play a role in the composition of this sample. 
The intrinsic luminosity densities, $L_{\nu i}(eq)$,
are plotted versus the
rest frame frequency $\nu_{i}(eq)$ in
Figure \ref{fig:lumeq}, and $L_{i}(eq)$ versus $\nu_{i}(eq)$ is plotted
in Figure \ref{fig:luminteq}. By 
fitting all the data plotted in Figures \ref{fig:lumeq} and 
\ref{fig:luminteq}, one finds that $L_{\nu i}(eq)\propto \nu_i(eq)^{2.3\pm0.1}$
with a reduced $\chi^2$ equal to 4.6 and a correlation coefficient $r=0.91$,
and $L_{i}(eq)\propto \nu_i(eq)^{3.3\pm0.1}$ with a reduced $\chi^2$ equal 
to 4.0 and a correlation coefficient $r=0.96$.  Both of these power law
fits are very significant since $r$ is close to unity.  Figures 
\ref{fig:lumz} and
\ref{fig:lumintz} show $L_{\nu i}(eq)$ and $L_{i}(eq)$ respectively
as functions of $(1+z)$; fitting all these data, one finds that 
$L_{\nu i}(eq)\propto(1+z)^{2.5\pm1.6}$ with a reduced $\chi^2$ equal to 1.3
and a correlation coefficient $r=0.16$, and 
$L_{i}(eq)\propto (1+z)^{0.9\pm2.2}$ with a reduced $\chi^2$ equal to 
1.4 and a correlation coefficient $r=0.04$.  
Similar results are found for $L_{\nu i}(IC)$ and $L_{i}(IC)$.

Figures \ref{fig:lumeq} \& \ref{fig:luminteq} show a trend in
$L_{\nu i}(eq)$ and $L_{i}(eq)$ with $\nu_{i}(eq)$
that can be explained 
by the strong dependence of these quantities on the Doppler
factor and a systematic effect introduced by
assuming the observed frequency is the peak frequency.  Equations 
(\ref{eq:inlum}) \& (\ref{eq:deq}) indicate
that $L_{\nu i}(eq)\propto{\delta_{eq}}^{-3}\propto \nu_{op}^{6.9}$
and that $\nu_i(eq)\propto\nu_{op}{\delta_{eq}}^{-1}\propto \nu_{op}^{3.3}$,
for $\alpha=0.75$.  
Since it is assumed that
$\nu_{op}=\nu_{o}$ when computing $\delta_{eq}$,
the offset of $\nu_{o}/\nu_{op}$ from unity will cause $L_{\nu i}(eq)$
to be offset by a factor of $(\nu_{o}/\nu_{op})^{6.9}$ and $\nu_i(eq)$
to be offset by a factor of $(\nu_{o}/\nu_{op})^{3.3}$ from their actual
values.  One would expect to
find a systematic effect that would increase the range of $L_{\nu
i}(eq)$ and $\nu_i(eq)$ such that $L_{\nu i}(eq)\propto\nu_i(eq)^{2.1}$.
In fact, one obtains $L_{\nu i}(eq)\propto \nu_i(eq)^{2.3\pm0.1}$ by
fitting the data in Figure \ref{fig:lumeq}.  A similar calculation
gives the prediction that $L_{i}(eq)\propto\nu_i(eq)^{3.1}$, and the
fits of the data in Figure \ref{fig:luminteq} give $L_{i}
(eq)\propto \nu_i(eq)^{3.3\pm0.1}$.  The spread over such a large range
of values for $L_{\nu i}(eq)$, $L_{i}(eq)$, and $\nu_{i}(eq)$ can thus be
explained by this systematic effect.  On the other hand, 
the errors in Figures
\ref{fig:lumeq} \& \ref{fig:luminteq} are quite large, and 
all the $L_{\nu i}(eq)$ and $L_{i}(eq)$
are still consistent with each other.

It is interesting to note that
$I_{\nu}(eq)\propto\nu_{i}(eq)^{2.2\pm0.1}$ and
$I_{\nu}(IC)\propto\nu_{i}(IC)^{2.0\pm0.1}$ for this sample, 
where $I_{\nu}\propto  
L_{\nu i}/{\theta_d}^2$.  This dependence would be expected if 
these sources were intrinsically similar, have black a body spectrum, and 
are observed in the
Rayleigh-Jeans size of the black body
spectrum.  It should be noted that the small dispersion in intrinsic brightness
temperatures described above (\S \ref{ssec:GPCM}) and the fact that 
$I_{\nu}$ is approximately proportional to $\nu^2$ go hand in hand.

Table \ref{tb:lum} lists the mean and
standard deviation of the mean for the logarithmic values of $L_{\nu
o}$, $L_{o}$, $L_{\nu i}(eq)$, $L_{i}(eq)$, $L_{\nu i}(IC)$, and
$L_{i}(IC)$, for each class of AGN.  Comparing $L_{\nu i}(eq)$ to
$L_{\nu i}(IC)$ and $L_{i}(eq)$ to $L_{i}(IC)$,
agreement
for each class of source is found within about $2\sigma$.
Ranges of
intrinsic luminosity densities and intrinsic luminosities
for all classes of AGN are $10^{32.4\;
to\;33.1}\mbox{ergs s}^{-1}\mbox{Hz}^{-1}$ and $10^{42\;
to\;43}
\mbox{ergs s}^{-1}$.

\section{DISENTANGLING RELATIVISTIC EFFECTS IN RADIO-LOUD
AGN\label{sec:anres}}

The effects of relativistic motion can be described by two
quantities:
the bulk Lorentz factor of the outflow, $\Gamma$, and the angle
subtended by
the outflow direction and the line of sight to the observer, $\phi$.
Note that the bulk Lorentz factor is determined by the speed of the
outflow relative to the speed of light, $\beta$, by the familiar
equation $\Gamma=1/ \sqrt{1-\beta^2}$.
 
Two observable quantities
which are combinations of $\Gamma$ and $\phi$ are
the Doppler factor, $\delta$, and the apparent
speed projected onto the plane
of the sky relative to the speed of light, $\beta_{app}$.
These are expressed in terms of $\beta$ and $\phi$ as
\begin{equation}
\delta={\sqrt{1-\beta^2}\over 1-\beta\cos\phi},
\label{eq:delta}
\end{equation}
and
\begin{equation}
\beta_{app}={\beta\sin\phi\over1-\beta\cos\phi}.
\label{eq:betapp}
\end{equation}
Equations (\ref{eq:delta}) and (\ref{eq:betapp})
can be used to solve for $\Gamma$ and $\phi$ in terms of
$\delta$ and $\beta_{app}$, and it can be shown that
\begin{equation}
\Gamma={\beta_{app}^2+\delta^2+1\over 2\delta}
\label{eq:gamma}
\end{equation}   
and
\begin{equation}
\tan\phi={2\beta_{app}\over \beta_{app}^2+\delta^2-1}
\label{eq:phi}
\end{equation}
(see GPCM93; Daly, Guerra, \& G\"uijosa 1996).
Thus, $\Gamma$ and $\phi$ can be estimated
for relativistic outflows in AGN, by combining
observational estimates of $\delta$ and $\beta_{app}$.
Here $\delta_{eq}$ and $\delta_{IC}$ are used to estimate $\delta$.

\subsection{Apparent Bulk Flow Versus Pattern Flow \label{ssec:pat}}

It has been suggested that the pattern velocities, the velocities of
radio
features observed in VLBI proper motion studies, are
greater than the bulk velocities, the velocities
responsible for the observed Doppler boosting
(see, for example, Lind \& Blandford 1985).
This issue was addressed by GPCM93, who conclude that the data are
consistent
with pattern velocities equaling bulk velocities for the sources in
their
sample with observed superluminal motion.

Vermeulen \& Cohen (1994, hereafter VC94) examine
the equivalence of pattern velocities and bulk velocities in
detail using Monte Carlo simulations of a simple model where
the angle between the direction of outflow and the line
of sight is allowed to vary randomly, but the bulk Lorentz factor
is the same in every jet.  They find that the pattern
Lorentz factor, $\Gamma_p$, would have to be
about twice the bulk Lorentz factor, $\Gamma_b$,
in order to be consistent with the data from the GPCM93 sample.  They
point
out, however, that for a broad
distribution of bulk Lorentz factors the data are easily consistent
with
the $\Gamma_p\simeq\Gamma_b$.

It is assumed, here and throughout, that $\Gamma_p\simeq\Gamma_b$, and any
departures from this equality are of order unity.

\section{ESTIMATES OF BULK LORENTZ FACTORS AND VIEWING
ANGLES\label{sec:sol}}

\subsection{A Subsample with $\beta_{app}$ Compiled\label{ssec:VC}}

VC94 compile a list of 66 radio sources with
multi-epoch VLBI observations of the internal proper motions.  Table 1 in VC94
lists $\beta_{app}$ for the features in these radio sources.  There are 43
sources that overlap the GPCM93 and the VC94 samples, and these sources are
examined below in order to estimate $\Gamma$ and $\phi$ 
(\S \ref{ssec:solve}); preliminary results were presented by Daly, Guerra,
\& G\"uijosa (1996).

Table \ref{tb:in}
lists the 43 sources that overlap the GPCM93 and VC94 samples.  Column
(1) lists the IAU Name, column (2) lists the common name if it exists,
Column (3) gives the redshift listed by VC94, Column (4) lists the 
$\beta_{app}$ estimates from VC94, except for 0108+388 where the
estimate is based on more recent observations by Taylor, Readhead, \& Pearson
(1996).
When multiple components 
are listed by VC94, the weighted mean is taken for $\beta_{app}$.
Figure \ref{fig:betapp} shows $\beta_{app}$ versus $(1+z)$ for the 43 sources 
in the overlapping sample. 
Columns (5) and
(6) give estimates of $\delta_{eq}$ and $\delta_{IC}$, respectively,
which are discussed in \S \ref{ssec:GPCM} above.

\subsection{Values for $\Gamma$ and $\phi$\label{ssec:solve}}

The values of $\beta_{app}$,  $\delta_{eq}$, and $\delta_{IC}$ 
listed in Table \ref{tb:in} can be used with equations (\ref{eq:gamma}) 
and (\ref{eq:phi}) to produce two sets of estimates for $\Gamma$ and 
$\phi$.   In the cases where only bounds on $\beta_{app}$ or $\delta$
are available, bounds on 
$\Gamma$ and $\phi$ are computed.

Table \ref{tb:out} lists the solutions for $\Gamma$ and $\phi$ using the
equipartition Doppler factor, called $\Gamma_{eq}$ and
$\phi_{eq}$, and using the inverse Compton Doppler factor, called
$\Gamma_{IC}$ and $\phi_{IC}$.  Columns (1) and (2) list the IAU name and
the common name respectively, Column (3) gives the redshift listed by
VC94, Columns (4) and (5) list $\Gamma_{eq}$ and $\phi_{eq}$
respectively, and columns (6) and (7) list $\Gamma_{IC}$ and $\phi_{IC}$
respectively.  Both sets of solutions for $\Gamma$ and $\phi$ agree
within errors
for any given source in Table \ref{tb:out}.

Table \ref{tb:med} lists the median values of
$\Gamma_{eq}$, $\phi_{eq}$, $\Gamma_{IC}$, and $\phi_{IC}$ for each
class of source, excluding those values which are upper or lower bounds.
The median
values of the equipartition set ($\Gamma_{eq}$ and $\phi_{eq}$) and the
inverse Compton set ($\Gamma_{IC}$ and $\phi_{IC}$) of estimates agree
well, except for the median $\Gamma_{eq}$ and $\Gamma_{IC}$
of the LDQs which are a factor of about two greater in the equipartition
case.  Note that excluding 3C216 from the LDQs would not significantly
change the
median values for this category.

Figures \ref{fig:gamtheeq}a \& \ref{fig:gamtheeq}b show $\Gamma_{eq}$
versus
$\phi_{eq}$ with and without errors respectively,
and Figures \ref{fig:gamtheic}a \& \ref{fig:gamtheic}b show $\Gamma_{IC}$
versus $\phi_{IC}$ with and without errors respectively.
The figures without errors are
shown so that the symbols are more visible to the reader and
bounds on $\Gamma$ and $\phi$ are denoted by arrows in these figures.
Figures \ref{fig:zeq}a \& \ref{fig:zeq}b show $\phi_{eq}$ and $\Gamma_{eq}$ as
functions of $(1+z)$ respectively, 
and Figures \ref{fig:zic}a \& \ref{fig:zic}b show $\phi_{IC}$ and
$\Gamma_{IC}$ as functions of $(1+z)$ respectively.

The median values for both sets of estimates of $\Gamma$ and $\phi$
make it possible to compare different classes of AGN in the
context of
orientation unified models (see discussion in \S \ref{ssec:comm}).
A detailed discussion of each set of $\Gamma$ and $\phi$ estimates
follows in \S\S \ref{ssec:soleq} \& \ref{ssec:solic}.

\subsubsection{Solutions Using the Equipartition Doppler
Factor \label{ssec:soleq}}

The eight BL Lacs studied here span a relatively large range of $\Gamma_{eq}$;
one source has a value of $\Gamma_{eq}=1.4\pm0.2$, and two sources
have $\Gamma_{eq}$ between 40 and 100, although these two sources
have large errors on $\Gamma_{eq}$.
The remaining five BL Lacs 
have $\Gamma_{eq}$ between 3 and 6.
The values of $\phi_{eq}$ for BL Lacs span a range from about 
$0^{\circ}$ to
about $50^{\circ}$, with six of the eight
BL Lacs having $\phi_{eq}\les20^{\circ}$.

Seven of the
eight CDHPQs have $\phi_{eq}\les12^{\circ}$, and at least
five of these sources
have $\phi_{eq}\les6^{\circ}$.
Three out of eight CDHPQs have
$\Gamma_{eq}$ between 3 and 10, and two
CDHPQs have 
$\Gamma_{eq}$ between 20 and 40.  Three sources have limits on
$\Gamma_{eq}$ and $\phi_{eq}$ instead of values (see \S
\ref{ssec:GPCM}).  Two of these sources with bounds, 1156+295
(4C 29.45) and 2230+114 (CTA 102), 
stand out in Figures \ref{fig:gamtheeq}a,b. 
The bounds
on $\Gamma_{eq}$ for these two sources, $\Gamma_{eq}<405$ for 1156+295
and $\Gamma_{eq}>2.8$ for 2230+114, still allow these sources to 
have Lorentz factor similar to other CDHPQs.  

Five CDLPQs have
$\Gamma_{eq}$ between 5 and 10, and
three have $\Gamma_{eq}$ between 15 and 30.  The other two sources
only have bounds on $\Gamma_{eq}$.  Eight out of ten
CDLPQs have
$\phi_{eq}\les14^{\circ}$.

The three CDQ-NPIs all have $\Gamma_{eq}<9$, and
have quite different values for $\phi_{eq}$.  The sources in this small 
category in reality belong to the CDHPQs or CDLPQs, and may contain
both types. 

Five LDQs have $\Gamma_{eq}$ between 10 and 50.  These
values of $\Gamma_{eq}$ are larger than the typical values
for the other classes of AGN studied
here.  One source has $\Gamma_{eq}$ less than 10 (4C 21.35,
$\Gamma_{eq}=2.8\pm2.9$), while another has a lower bound
$\Gamma_{eq}>3$. 
Only one LDQ has $\phi_{eq}$ less than
$15^{\circ}$ (3C216), while the other six LDQs have
$\phi_{eq}$ between $15^{\circ}$ and $41^{\circ}$.  
LDQs have the larger values for $\phi_{eq}$ than most of the BL Lacs,
CDQs (HP, LP, and NPI) studied here. 

RGs have the largest values and widest range of
$\phi_{eq}$ of all the categories.  Four out of seven RGs have
$45^{\circ}<\phi_{eq}<135^{\circ}$,
which suggests that these sources have
outflows that typically lie close to the plane of the sky (see \S
\ref{ssec:comm}).  RGs also have the lowest $\Gamma_{eq}$ of all the
categories, with at least four RGs having $\Gamma_{eq}<2$.  Note that
the radio galaxies in this sample are
compact steep spectrum sources or FRI; there are 
no FRII radio galaxies in this sample.

\subsubsection{Solutions Using the Inverse Compton Doppler
Factor\label{ssec:solic}}

The results for $\Gamma_{IC}$ and $\phi_{IC}$ are quite similar to those
for $\Gamma_{eq}$ and $\phi_{eq}$, and agree well within errors.  This
agreement is not surprising considering the correlation and 
agreement found between
$\delta_{IC}$ and $\delta_{eq}$ (see \S \ref{ssec:GPCM}).  

Five out of eight BL Lacs have
$\Gamma_{IC}$ between 3 and 5, while the total range
for all BL Lacs is between 1.5 and 60.
Seven BL Lacs have $\phi_{IC}<20^{\circ}$.  
Five out of eight CDHPQs have $\Gamma_{IC}$
between 8 and 16, while the other three have various
upper and lower bounds.
These same five CDHPQs all have
$\phi_{IC}<10^{\circ}$. 
Six out of ten CDLPQs have $\Gamma_{IC}$ between 3 and 11,
and two have $\Gamma_{IC}$ around 20.  Eight CDLPQs have
$\phi_{IC}<14^{\circ}$.
Five out of seven LDQs have $\Gamma_{IC}$ between 10 and
30, and five LDQs have $\phi_{IC}$ between $15^{\circ}$
and $45^{\circ}$.  As in the equipartition case only one LDQ has 
$\phi_{IC}$ less than $15^{\circ}$.
Three out of seven RGs have $\phi_{IC}$ between
$45^{\circ}$ and $90^{\circ}$, and three RGs have
$\phi_{IC}>90^{\circ}$.  At least six RGs have $\Gamma_{eq}<2$.

\subsection{Implications for Orientation
Unified Models of AGN
\label{ssec:comm}}

The results presented \S \ref{ssec:solve} are consistent with an
orientation unification scheme for radio-loud AGN (e.g.
Antonucci 1993).
Common orientation models have 
a torus or disk of material that absorbs or obscures
the broad absorption lines and
other non-stellar radiation
emitted from the
nuclear region of the AGN, and the classification as a radio galaxy occurs
if the torus intersects the line of sight from the nuclear region to the
observer.  The radio jet axis in radio-loud AGN is thought to be more or
less perpendicular to the plane of the torus or disk. 
In such models (e.g. Barthel 1989), RGs, LDQs, and CDQs, are 
viewed with different
angles with respect to the jet axis ($\phi$, which is referred 
to as the viewing angle): radio galaxies have the
largest angles ($\phi\ges45^{\circ}$), LDQs have smaller angles
($20^{\circ}\les\phi\les45^{\circ}$), and CDQs have the smallest angles
($\phi\les20^{\circ}$).  

The RGs, LDQs, and CDQs in overlap of the GPCM93 and VC94 samples have
estimates of $\phi$ consistent with this orientation unification
scheme.
The RGs in this sample have a median $\phi_{eq}$ of $110^{\circ}$ and a
median $\phi_{IC}$ of $81^{\circ}$, which suggests that they have 
jets that lie close
to the plane of the sky.
LDQs typically have a median $\phi_{eq}$ of $26^{\circ}$
and a median $\phi_{IC}$
of $25^{\circ}$.
CDLPQs have a median $\phi_{eq}$ of $7^{\circ}$
and a median $\phi_{IC}$ of $6^{\circ}$, while CDHPQs have a median
$\phi_{eq}$ of $4^{\circ}$
and a median $\phi_{IC}$ of $3^{\circ}$.  
The median values for CDQ-NPIs are insignificant since only two sources
are used to compute them.
It is interesting to note that CDHPQs and CDLPQs have similar median
$\phi_{eq}$ and $\phi_{IC}$.  This suggests that it is intrinsic
differences between these two classes of AGN (not 
viewing angles) that determine the amount of optical polarization observed.

The BL Lacs in this sample cover a wide range of
$\Gamma$ and $\phi$, but have viewing angles typically
less than
$45^{\circ}$.   
An extension of the orientation model described above assigns BL Lacs to
the parent population of FRI (edge-darkened, \cite{fr}) 
RGs, where BL Lacs are seen
at smaller viewing angles and radio selected BL Lacs have smaller angles
than X-ray selected BL Lacs (see Padovani \& Urry 1992).
The sample of BL Lacs examined
here are
almost all radio selected.  Mkn 421 (1101+384) is the only X-ray
selected BL Lac in this sample, but
does not stand out against the other BL Lacs in terms of its $\Gamma$ or
$\phi$ estimates.  

The BL Lacs in this sample
have a median $\phi_{eq}$ of $12^{\circ}$ 
and a median 
$\phi_{IC}$ of $14^{\circ}$.  These results are consistent with the 
those found by
Kollgaard et al. (1996), who estimate the viewing angles for samples of
radio selected BL Lacs, X-ray selected BL Lacs, and RGs by comparing the
core enhancement relative to the more diffuse radio emission for each
class of AGN.  Assuming a constant bulk Lorentz factor,
they find that their samples of
radio selected BL Lacs, X-ray selected BL
Lacs, and RGs 
have average viewing angles of
$10^{\circ}$, $20^{\circ}$, and $60^{\circ}$
respectively.  

CDQs in this sample are centered around a $\Gamma\approx9$.  If it is
assumed that the distribution of bulk Lorentz factors for all classes of
radio-loud AGN is centered around $\Gamma\approx9$, 
then one might naively expect the mean or median $\Gamma$ for each of
class of AGN to similar.  Selection biases likely
play a role in the composition of the samples studied, and
systematic effects can affect the values that can be obtained from the data.

LDQs have median $\Gamma_{eq}$ and $\Gamma_{IC}$ that are larger than
those for the CDQs, and these two estimates are different almost by a
factor of two, yet the mean ratio of $R=\delta_{eq}/\delta_{IC}$ is only
about 0.76 for LDQs.  This could be caused by the 
sensitivity of the estimates of
$\Gamma$ to the precise values
of $\beta_{app}$ and $\delta$ when these to parameters are close to one
(see eq. \ref{eq:gamma}), which is the case for this sample of LDQs.
A decrease of $\delta$ away from one will cause the corresponding
$\Gamma$
estimate to increase.  If the $\delta$ estimates of LDQs in this sample
are systematically
underestimated, then a noticeable systematic increase in the $\Gamma$ will
result, which could be caused by an overestimate of the peak
frequency (see eqs. \ref{eq:deq} and \ref{eq:dic}).  It is not clear
whether this is one of the causes of the larger $\Gamma$ for LDQs.

The lower $\Gamma$ for RGs compared other classes of AGN can be
explained as follows.  RGs will appear in this sample if
there cores are bright enough to have VLBI core size data, which occurs
if the sources are nearby (like M87 and NGC 6251) or have very
bright compact cores.  In particular, FRII radio galaxies are missing
from this sample since a majority of their cores are not significantly
Doppler boosted and they make difficult targets for VLBI.
In the unification scheme described above, RGs have viewing angles
around $90^{\circ}$.  Approximating $\phi\approx90^{\circ}$, one finds
that $\delta\approx\Gamma^{-1}$ from equation (\ref{eq:delta}).  Thus,
$\Gamma\approx\delta^{-1}$, and
those RGs with smaller $\Gamma$ 
will appear in this
sample since these RGs will have larger $\delta$ and brighter cores.  Even if 
the true distribution of
$\Gamma$ for RGs peaked at around 10, we may only be able to observe
RGs with $\Gamma$ from 1 to 2.  It should be noted that
RGs in this sample have values of
$\Gamma$ near unity which are 
in agreement with other evidence that Doppler boosting
is not significant in these sources (e.g. Taylor, Readhead, \& Pearson 1996).

There are two sources that seem to have 
viewing angles greater than $90^{\circ}$ based on equipartition and inverse
Compton Doppler factors.  One, 2352+495, has a lower bound on the viewing
angle, the other, 0710+439, is greater than $90^{\circ}$ at around $3\sigma$.
This would suggest that some RGs have observed outflows that 
are moving away from the observer, and that if there is an outflow moving
toward the observer it must be intrinsically fainter.

BL Lacs have lower values of $\Gamma$ than the CDQs, with a median
$\Gamma_{eq}$ of 4.2 and a median $\Gamma_{IC}$ of 5.  Since the BL Lacs
studied in this paper are almost exclusively radio selected, it would be
interesting to compare $\Gamma$ and $\phi$ estimates for a sample of
X-ray selected BL Lacs.  Kollgaard et al. (1996) find that the core
enhancements they observe for BL Lacs and RGs, require that $\Gamma>4.5$,
assuming a constant bulk Lorentz factor
for all these sources.  The estimates of $\Gamma$ found here are barely
consistent with $\Gamma>4.5$,
but a distribution of Lorentz factors may loosen the constraints placed
by Kollgaard et al. (1996).

\section{DISCUSSION\label{sec:disc}}

The two estimates of the Doppler factor for the sample
of radio-loud AGN 
discussed in \S\ref{ssec:GPCM} agree with
each other on average and provide a means of estimating the intrinsic
properties of these sources (e.g. brightness temperatures, luminosities)
and the parameters that describe the kinematics of the relativistic bulk
flow of the radio emitting plasma.  Although the results here are based
on rough
estimates of the Doppler factor, the average or typical
properties of 
different classes of AGN can be compared.

Care should be taken when using
the estimates of Doppler factors and
intrinsic properties computed here
since systematic errors contribute most of the
uncertainty. These systematic errors arise
from the simplification that the observed frequencies correspond to the peak
of the core emission.  Another concern with this sample is completeness.
The sample examined here was 
defined by GPCM93 as those which had VLBI core size data
in the literature at that point in time.  The sources that enter this sample
will have VLBI cores with
higher surface brightness than those excluded from the sample;
this will favor the brighter (more highly Doppler boosted) and more compact
(smaller angular size)
cores.  GPCM93 warn that this sample
is not complete in any statistical sense, and that biases may have a large
effect on the content of this sample.  

Sources in the first and second
Caltech-Jodrell Bank VLBI surveys (CJ1 and CJ2) form a complete flux-limited
sample of
almost 400 compact radio sources with VLBI measured core sizes 
(see Henstock et al. 1995, Xu et al. 1995) and some multi-frequency data. 
Doppler factor estimates for sources from these samples
are currently under investigation, including the use
of spectra to estimate the peak frequency of 
core components in order to more accurately compute Doppler
factors (\cite{prep}).  Large 
samples such as these 
could be used to estimate the Doppler factors and intrinsic properties
of different classes of AGN with greater confidence and statistical
weight. 

The intrinsic brightness temperatures for this 
sample are estimated using two sets of Doppler factor estimates,
$\delta_{eq}$ and $\delta_{IC}$, and
both histograms of these values are centered around $7.5\times10^{10}\mbox{
K}$ (see \S \ref{ssec:GPCM}).  
The intrinsic brightness temperatures are
similar for different classes of AGN, despite the fact that the 
observed brightness
temperatures are different for different classes of AGN.
The center of the distribution of intrinsic brightness temperatures
is a factor of 4
lower than what would be expected if the ``inverse Compton
catastrophe'' played a major role in the observed brightness temperature
cutoff, as noted previously by Readhead (1994).  This suggests that either a 
factor of about 4 has not been accounted for in these estimates (e.g. a 
geometric factor), or that a physical process other than the ``inverse Compton
catastrophe'' is limiting the 
intrinsic brightness temperature. 

Estimates of the intrinsic 
luminosity density and luminosity depend strongly on the Doppler factor
($L_{\nu i}\propto\delta^{-3}$ and $L_{i}\propto\delta^{-4}$), and systematic
errors are quite large for these estimates.  
The large scatter of $L_{\nu i}$ and $L_{i}$ is a cause for concern
since it can be attributed to systematic errors introduced by assuming
the observed frequencies 
correspond to the actual peak frequencies 
of these sources.  In any event, these
estimates can be compared to the Eddington luminosity for a supposed
massive central object.  The Eddington luminosity is 
$L_E=1.3\times10^{44}M_6\mbox{ ergs s}^{-1}$ where $M_6$ is the mass of
the central object in units of $10^6 M_{\odot}$ which gives a
range from $10^{44}\mbox{ to }10^{46}\mbox{ ergs s}^{-1}$ 
for central objects with $M_6=1
\mbox{ to }100$.  The mean intrinsic luminosity for the sources examined here
is around $10^{42}\mbox{ to }10^{43}\mbox{ ergs s}^{-1}$, while
only five objects (3 BL Lacs, 1 LDQ, and 1 RG)
have intrinsic luminosities greater than $10^{46}\mbox{ ergs s}^{-1}$
by more than $1\sigma$.  This is consistent with the radio 
luminosity produced being 0.1 to 1 percent of the Eddington luminosity for 
a central compact object with mass $10^6\mbox{ to }10^8 M_{\odot}$.

A subsample of 43 sources have observed proper motions in
the plane of the sky and were used by VC94 and others to
compute $\beta_{app}$.  The
values of $\beta_{app}$ for each of these sources
can be combined with the Doppler factor
estimates to solve for the bulk Lorentz factor, $\Gamma$, and the viewing
angle, $\phi$ (see \S
\ref{sec:anres}).  The estimates using the data compiled here
are consistent with the orientations
unification scheme for AGN (discussed in \S \ref{ssec:comm}),
where CDQs have the smallest $\phi$ ($<15^{\circ}$), 
LDQs have larger $\phi$ ($15^{\circ}<\phi<45^{\circ}$), and
RGs have the largest
$\phi$ ($>45^{\circ}$); note that the RGs in this sample are compact steep
spectrum sources or FRI.

Selection biases and systematic errors may play a role in these
estimates of $\Gamma$ and $\phi$, in particular for LDQs and RGs
(see \S \ref{ssec:comm}).
The RGs in this sample have lower
$\Gamma$ than the other classes of AGN, which is 
consistent with the parent population having
outflows that lie close to the plane of the sky (i.e. $\phi\sim
90^{\circ}$) and the flux-limited nature of the sample studied.  The 
LDQs are in a regime where small systematic errors in
$\delta$ can strongly affect $\Gamma$ estimates.

The estimates of $\phi$ and $\Gamma$ for the BL Lacs in this sample are
consistent with the results of Kollgaard et al. (1996) for radio selected
BL Lacs (see \S \ref{ssec:comm}).
Their estimates of $\phi$ are based on the core enhancement of radio
selected BL Lacs, X-ray selected BL Lacs, and RGs.
A sample of
X-ray selected BL Lacs with
estimates of $\phi$ and $\Gamma$ computed as in \S \ref{ssec:solve} would
give a consistency check for
both methods.

Marscher (1977)
discusses the effect of gradients in magnetic field and
density with radius on the relationships between the observed quantities
and the important physical parameters in a nonuniform source.
The effect on $\delta_{eq}$ and $\delta_{IC}$ of including 
inhomogeneities is constant
multiplicative factor from 0.7 to 1.4, and
the the functional dependence on observable quantities is 
identical to the homogeneous case; the form of the inhomogeneity
and the parameter choices are discussed by Marscher (1977).  It should be noted
that the intrinsic brightness temperatures will be changed by
the same constant factor, and that inhomogeneities do not account
for the factor of 4 in brightness temperature
that would allow the inverse Compton catastrophe to play a major role in
the brightness temperature distribution (\S \ref{ssec:GPCM}).

The spherical model for the radio cores of AGN, which is used throughout
this paper to compute Doppler factors, may not be as
realistic as those models which consider a jet-like geometry.
The expression
for $\delta_{IC}$ in the jet-like case, which is derived by GPCM93, is
\begin{equation}
\delta_{IC}(jet)=\left[\delta_{IC}(sph)\right]^{(4+2\alpha)/(3+2\alpha)},
\end{equation}
where $\delta_{IC}(sph)$ is the expression given by equation
(\ref{eq:dic}).  For a value 
of $\alpha=0.75$, $\delta_{IC}(jet)=\delta_{IC}(sph)^{1.2}$ which
is a power law only slightly greater than unity.  
The expression for $\delta_{eq}$ in the jet-like case is
\begin{equation}
\delta_{eq}(jet)=\left[\delta_{eq}(sph)\right]^{(13+2\alpha)/(9+2\alpha)}
(\sin\phi)^{2/(9+2\alpha)}
\left({{\theta_a}\over{\theta_b}}\right)^{1/(9+2\alpha)},
\label{eq:jet}
\end{equation}
where $\delta_{eq}(sph)$ is given by equation (\ref{eq:deq}),
$\phi$ is the angle subtended by the direction of the outflow and the line
of sight, and $(\theta_a/\theta_b)$ is the ratio of the angular sizes
of the  major and minor axes (see Appendix for the derivation
of this equation).  For $\alpha=0.75$,
$\delta_{eq}(jet)\propto (\sin \phi)^{0.19}$, and $\delta_{eq}(jet)\propto 
(\theta_a/\theta_b)^{0.09}$,
which are very weak dependences.  The relation between the
jet-like and spherical cases of the equipartition Doppler factor is 
$\delta_{eq}(jet)\propto\delta_{eq}(sph)^{1.4}$ which is
a power-law
close to unity.
The expressions for inverse Compton and equipartition
Doppler factors in the jet-like case appear to be
quite similar to those in the spherical case.  Thus, approximating a
spherical geometry for simplicity can still
provide useful Doppler factor estimates even if the true geometry of 
these outflows
is more jet-like.

\acknowledgments
The authors would like to thank Tom Herbig,
Alan Marscher, Tony Readhead, Larry Rudnick, 
David Schramm, Steve Unwin, Lin Wan, and
Greg Wellman for
helpful discussions.
This work was supported in part by the US National Science
Foundation, by an NSF Graduate
Fellowship, the Independent College Fund
of New Jersey, and by a grant from W. M. Wheeler III.

\appendix
\section{Relevant Formulae}

For the simple case
of a uniform sphere where the synchrotron-emitting particles have a power-law
energy distribution and move through a tangled homogeneous magnetic field, the 
Doppler factor needed to reconcile the predicted and observed X-ray fluxes 
(Marscher 1987) is
\begin{equation}
\delta_{IC}(sph)=f(\alpha)S_m \left[ {\ln \left(\nu_b / \nu_{op} \right)}
\over{S_x \theta_d^{6+4\alpha} \nu_x^{\alpha} \nu_{op}^{5+3\alpha}}
\right]^{1/(4+2\alpha)} (1+z),
\label{eq:dic}
\end{equation}
where $z$ is the redshift, 
$S_x$ is the observed X-ray flux density (in Jy) at frequency $\nu_x$
(keV), $\theta_d$ is the angular diameter of the sources (in mas), $\nu_b$
is the synchrotron high-frequency cutoff (assumed to be $10^5$ GHz), 
$\alpha$ is the optically thin spectral index (where the radio 
flux density 
$S_{\nu}\propto \nu^{-\alpha}$), $\nu_{op}$ is the observed frequency of 
the radio peak (in GHz), and $f(\alpha)=0.08\alpha+0.14\,$. The radio
flux density 
in equation (\ref{eq:dic}), $S_m$ (Jy), is the value obtained by extrapolating 
the optically thin flux density to the observed peak at $\nu_{op}$.
In equation (\ref{eq:dic}) and throughout this work, 
$\theta_d=1.8 \theta_{FWHM}$ and $\theta_{FWHM}=\sqrt{
\theta_a \theta_b}$ from VLBI observations, where $\theta_a$ and $\theta_b$
are the angular sizes of the major and minor axes respectively
(see G\"uijosa \& Daly 1996).

The equipartition Doppler factor is estimated by assuming the
energy density in magnetic field equals the energy 
density in synchrotron-emitting relativistic particles (Readhead 1994). 
In the notation used here,
\begin{equation}
\delta_{eq}(sph)=\bigg( \Big[10^3 F(\alpha)\Big]^{34}\Big[4h/y(z)\Big]^2
(1+z)^{15+2\alpha} S_{op}^{16} \theta_d^{-34}\nu_{op}^{2\alpha-35}
\bigg)^{1/(13+2\alpha)}~,
\label{eq:deq}
\end{equation}
where $S_{op}$ is the observed peak flux density and $\nu_{op}$ is
the observed
frequency of the peak.
The solutions for $F(\alpha)$ are given by Scott \&
Readhead
(1977), and the solution of interest here is $F(0.75)=3.4$ since
$\alpha=0.75$ is assumed throughout.
The function $y(z)=H_oa_or(z)/c$ (Peebles 1993) is a dimensionless
function of $\Omega_o$, $\Omega_{\Lambda}$, and
$z$, and contains the dependence
on the
coordinate distance to the source.
In equation
(\ref{eq:deq}), the Hubble's constant is parameterized as
$H_o=100~h~\hbox{km s}^{-1} \hbox{ kpc}^{-1}$. 
The dependence of $\delta_{eq}$ on cosmology is very weak.  For $\alpha=0.75$,
$\delta_{eq} \propto (a_or(z))^{-0.14} \propto (h/y(z))^{0.14}$
Throughout this paper, $\Omega_o=0.1$, $\Omega_{\Lambda}=0$, and $h=0.75$ are
assumed, where the functional form of $y(z)$ is 
that derived from equation (13.36) of Peebles (1993).
Note that the results of G\"uijosa \& Daly (1996)
were obtained assuming an Einstein-de Sitter cosmology ($\Omega_o=1.0$, 
$\Omega_{\Lambda}=0$) with $h=1$.

The expression for $\delta_{eq}(jet)$ (eq. \ref{eq:jet}) can be derived
by equating equipartition magnetic field, $B_{eq}$,
to the synchrotron self-absorbed
magnetic field, $B_{SSA}$, as is done when 
deriving $\delta_{eq}(sph)$ (Readhead 1994).
The equipartition magnetic field is given by
\begin{equation}
B_{eq}=(2\; C_{me}\; i_1(\alpha)\; {\tau}_p^{-1}\; [1-\exp(-{\tau}_p)]\; I_p 
\; {\nu_p}^{\alpha}\; l^{-1})^{2/7} 
\label{eq:beq}
\end{equation}
where $C_{me}$ and $i_1$ are defined by Leahy (1991), $\tau_p$ is the 
optical depth of the peak of the spectrum (see Scott \& Readhead 1977),
$\nu_p$ is the 
frequency of the peak as measured in the frame
of the synchrotron-emitting plasma, $I_p$ is the peak specific intensity
as measured in 
the plasma frame, $l$ is the line of sight depth as measured in 
the plasma frame, and $\alpha$ is the optically thin spectral index 
($I_{\nu}\propto {\nu}^{-\alpha}$).
The synchrotron self-absorbed magnetic field is given by
\begin{equation}
\quad B_{SSA}= \biggl({\pi \over 6}\; {c_1(\alpha) \over c_2(\alpha)}\;
{\tau}_p^2 \; [1-\exp(-{\tau}_p)]^{-1}\biggr)^2\;
 {I_p}^{-2}\; \nu_{p}^5
\label{eq:bssa}
\end{equation}
where $c_1(\alpha)$ and $c_2(\alpha)$ are tabulated by Marscher (1977).
To simplify these expression we define 
\begin{equation}
G_{eq}(\alpha)\equiv
(2\; C_{me}\; i_1(\alpha)\; {\tau}_p^{-1}\; [1-\exp(-{\tau}_p)])^{2/7}
\end{equation} 
so
that $B_{eq}=G_{eq}(\alpha) ( I_p 
\; {\nu_p}^{\alpha}\; l^{-1})^{2/7}$, and
\begin{equation}
G_{SSA}(\alpha)\equiv \biggl({\pi \over 6}\; {c_1(\alpha) \over c_2(\alpha)}\;
{\tau}_p^2 \; [1-\exp(-{\tau}_p)]^{-1}\biggr)^2
\end{equation}
so that $B_{SSA}=G_{SSA}(\alpha)\; {I_p}^{-2}\; \nu_{p}^5$. 

The equation for $\delta_{eq}(sph)$ is derived by setting 
$B_{eq}=B_{SSA}$,
$I_p= S_{op}\; (1+z)^3\; \delta^{-3}\; \theta_{d}^{-2}$, 
$\nu_p=\nu_{op}\; (1+z)\; \delta^{-1}$, $l=\theta_{d}\; a_o r(z)\; 
(1+z)^{-1}$,
and solving for $\delta$.  It should be noted that
a spherical geometry is assumed
when using these expressions for $I_p$, $\nu_p$, and $l$ (see GPCM93).
The result,
\begin{equation}
\delta_{eq}(sph)^{13+2\alpha}=
G_{eq}(\alpha)^7\; G_{SSA}(\alpha)^{-7}\; [a_o r(z)]^{-2}\;
(1+z)^{15+2\alpha}\; S_{op}^{16}\; \theta_{d}^{-34}\;
\nu_{op}^{2\alpha - 35},
\end{equation}
reduces to the equation of Readhead (1994) for the equipartition 
Doppler factor (see eq. \ref{eq:deq}).   

The equipartition Doppler factor for a jet-like geometry is derived by
using the appropriate expressions for 
this case, which are $I_p= S_{op}\; (1+z)^3\; 
\delta^{-2-[2\alpha/(5+2\alpha)]}\; \theta_{d}^{-2}$, $\nu_p=\nu_{op}\; 
(1+z)\; \delta^{-1+[2/(5+2\alpha)]}$, and $l=\sqrt{\theta_b/\theta_a}\;
\theta_{d}\; a_o r(z)\; 
(1+z)^{-1}\; \delta^{-1}\; (\sin \phi)^{-1}$, where
$\theta_a$ and $\theta_b$ are angular sizes of the major and minor axes 
respectively, and $\phi$ is the outflow angle
(see GPCM93).  The result using these substitutions is
\begin{equation}
\delta_{eq}(jet)^{9+2\alpha}= G_{eq}(\alpha)^7\; 
G_{SSA}(\alpha)^{-7}\; [a_o r(z)]^{-2}\;
(1+z)^{(15+2\alpha)}\; S_{op}^{16}\; \theta_{d}^{-34}\;
\nu_{op}^{(2\alpha - 35)}\; (\sin \phi)^{2}\; (\theta_a/\theta_b), 
\end{equation}
which reduces to equation (\ref{eq:jet}).

The brightness temperature, $T_B$, is a useful quantity to use when discussing
radiative processes.  The observed brightness
temperature at the peak is 
\begin{equation}
T_{Bo}=1.77\times10^{12}{{S_{op}}\over{\theta_d^{2}\nu_{op}^{2}}}
\end{equation}
(e.g. Readhead 1994).
The intrinsic brightness
temperature, $T_{Bi}$, can be related to the observed brightness
temperature for a relativistic moving sphere using $T_{Bi}=T_{Bo}(1+z)/\delta$,
where $z$ is the redshift to the source and $\delta$ is the Doppler
factor.  
These estimates are given by:
\begin{equation}
T_{Bi}=1.77\times10^{12}{{S_{op}}\over{\theta_d^{2}\nu_{op}^{2}}}{{(1+z)}
\over{\delta}}
\end{equation}
(e.g. Readhead 1994). Estimates of $T_{Bi}$ using 
$\delta_{eq}$ are denoted $T_{Bi}(eq)$, and
those using $\delta_{IC}$ are denoted $T_{Bi}(IC)$.

\clearpage

\begin{deluxetable}{cccc}
\footnotesize
\tablecaption{Mean Brightness Temperatures
\label{tb:bri}}
\tablewidth{0pt}
\tablehead{
\colhead{}  &   
\colhead{Mean} & \colhead{Mean} & \colhead{Mean} \\
\colhead{Class} & $T_{Bo}/10^{11}\mbox{ K}$ & 
$T_{Bi}(eq)/10^{11}\mbox{ K}$ &
$T_{Bi}(IC)/10^{11}\mbox{ K}$ 
}
\startdata
All Sources & $2.3\pm0.3$ &  $0.87\pm0.03$ &
 $0.73\pm0.03$  \\
BL Lacs     & $2.2\pm0.6$ &  $0.86\pm0.06$ &
 $0.75\pm0.06$  \\
CDHPQs	    & $3.5\pm0.6$ & $0.79\pm0.03$ &
 $0.75\pm0.06$  \\
CDLPQs      & $2.4\pm0.5$ & $0.89\pm0.05$ & 
 $0.80\pm0.09$  \\
CDQ-NPIs    & $0.74\pm0.24$  & $0.73\pm0.19$ &
 $0.51\pm0.13$  \\ 
LDQs        & $2.0\pm1.7$ &  $1.0\pm0.1$ &
$0.78\pm0.08$  \\
LDQs$-$3C216& $0.30\pm0.09$ & $1.1\pm0.1$ &
 $0.76\pm0.08$  \\
RGs         & $0.36\pm0.10$ & $0.97\pm0.13$ &
 $0.65\pm0.13$ 
\enddata
\end{deluxetable}

\begin{deluxetable}{ccccccc}
\footnotesize
\tablecaption{Mean $\delta_{eq}$, $\delta_{IC}$ for True and Reassigned Angular
Sizes \label{tb:test}}
\tablewidth{0pt}
\tablehead{
\colhead{Class} & \colhead{No.} &  \multicolumn{2}{c}{True} &
&  \multicolumn{2}{c}{Reassigned} \\
\cline{3-4} \cline{6-7} \\
\colhead{} & \colhead{} & \colhead{Mean $\delta_{eq}$}
& \colhead{Mean $\delta_{IC}$} & & \colhead{Mean $\delta_{eq}$}
& \colhead{Mean $\delta_{IC}$} 
}
\startdata
All Sources & 100 & $6.0\pm0.9$ & $5.1\pm0.6$ & & $26\pm14$ & $9.2\pm2.4$ \\
BL Lacs     & 27 & $4.5\pm1.2$ & $3.4\pm0.7$ & & $12\pm7$ & $5.8\pm2.7$ \\
CDHPQs	    & 24 & $9.0\pm1.5$ & $8.4\pm1.0$ & & $90\pm56$ & $23\pm10$ \\
CDLPQs      & 22 & $6.9\pm1.5$ & $6.2\pm1.4$ & & $21\pm12$ & $13\pm6$ \\
CDQ-NPIs    & 7 & $5.1\pm2.2$ & $5.3\pm1.8$ & & $26\pm18$ & $14\pm9$ \\
LDQs        & 11 & $6.1\pm5.5$ & $3.6\pm2.9$ & & $3.5\pm2.0$ & $2.0\pm0.9$ \\
LDQs$-$3C216& 10 & $0.56\pm0.22$ & $0.66\pm0.22$ & & $0.9\pm0.5$ & $0.6\pm0.2$ \\
RGs         & 9 & $0.55\pm0.17$ & $0.65\pm0.14$ & & $1.3\pm0.8$ & $1.1\pm0.5$ 
\enddata
\end{deluxetable}

\begin{deluxetable}{ccccccc}
\footnotesize
\tablecaption{Mean Luminosity Densities and Luminosities
\label{tb:lum}}
\tablewidth{0pt}
\tablehead{
\colhead{}  &   \colhead{Mean} & \colhead{Mean} & \colhead{Mean} &
\colhead{Mean} & \colhead{Mean} & \colhead{Mean} \\
\colhead{Class} & $\log L_{\nu o}$ & $\log L_{o}$ & $\log L_{\nu i}(eq)$ &
$\log L_{i}(eq)$ & $\log L_{\nu i}(IC)$ & $\log L_{i}(IC)$ \\
\colhead{}  &
$\left({{\mbox{ergs}}
\over{\mbox{s Hz}}}\right)$ &
$\left({{\mbox{ergs}}
\over{\mbox{s}}}\right)$ &
$\left({{\mbox{ergs}}
\over{\mbox{s Hz}}}\right)$ &
$\left({{\mbox{ergs}}
\over{\mbox{s}}}\right)$ &
$\left({{\mbox{ergs}}
\over{\mbox{s Hz}}}\right)$ &
$\left({{\mbox{ergs}}
\over{\mbox{s}}}\right)$ 
}
\startdata
All Sources & $33.6\pm0.1$ &  $43.3\pm0.1$ &
 $32.9\pm0.2$  & $42.7\pm0.3$ &  $32.6\pm0.2$ &
 $42.2\pm0.2$\\
BL Lacs     & $32.8\pm0.2$ &  $42.4\pm0.2$ &
 $32.8\pm0.5$ & $42.5\pm0.7$ &  $32.6\pm0.4$ &
 $42.3\pm0.5$ \\
CDHPQs	    & $34.3\pm0.1$ & $44.1\pm0.2$ &
 $32.0\pm0.3$ & $41.3\pm0.4$ &  $31.9\pm0.2$ &
 $41.1\pm0.3$ \\
CDLPQs      & $34.4\pm0.2$ & $44.2\pm0.3$ & 
 $33.0\pm0.5$ & $42.6\pm0.6$ &  $32.7\pm0.3$ &
 $42.2\pm0.4$ \\
CDQ-NPIs    & $34.0\pm0.3$  & $43.8\pm0.4$ &
 $32.8\pm0.7$ & $42.4\pm0.9$ &  $32.5\pm0.5$ &
 $42.0\pm0.7$ \\ 
LDQs        & $33.2\pm0.3$ &  $43.0\pm0.3$ &
$34.4\pm0.8$ & $44.8\pm1.1$ &  $34.0\pm0.7$ &
 $44.3\pm0.9$ \\
LDQs$-$3C216& $33.1\pm0.3$ & $42.9\pm0.3$ &
 $35.0\pm0.7$ & $45.6\pm0.9$ &  $34.5\pm0.5$ &
 $45.0\pm0.7$ \\
RGs         & $31.9\pm0.5$ & $41.7\pm0.5$ &
 $33.8\pm1.1$ & $44.3\pm1.3$ &  $33.1\pm0.7$ &
 $43.3\pm0.9$
\enddata
\end{deluxetable}

\begin{deluxetable}{cccccc}
\scriptsize
\tablecaption{The Overlap of the GPCM and VC Samples \label{tb:in}}
\tablewidth{0pt}
\tablehead{
\colhead{Source} & \colhead{Name} &  \colhead{$z$} & \colhead{$\beta_{app}$} &
\colhead{$\delta_{eq}$} & \colhead{$\delta_{IC}$}\\
\colhead{(1)} & \colhead{(2)} & \colhead{(3)} & \colhead{(4)} &
\colhead{(5)} & \colhead{(6)}
}
\startdata
\underline{BL Lacs}\\
0454+844 &... & 0.112 & $0.9\pm0.3$ & 1.3 & 2.4 \nl
0851+202 & OJ 287 & 0.306 & $4.0\pm0.4$ & 11 & 6.8 \nl
1101+384 & Mkn 421 & 0.031 & $2.50\pm0.04$ & $2.2$ & $0.92$ \nl
1308+326 &... & 0.996 & $18\pm9$ & 4.0 & 5.2 \nl
1749+701 &... & 0.770 & $10\pm1$ & 0.6 & 0.9 \nl
1803+784 &... & 0.684 & $0.1\pm0.9$ & 5.3 & 6.6 \nl
2007+776 &... & 0.342 & $3.4\pm0.7$ & 2.8 & 3.6 \nl
2200+420 & BL Lac & 0.069 & $4.6\pm0.2$ & 5.0 & 3.4 \nl
\underline{CDHPQ}\\
0212+735 &... & 2.370 & $8\pm4$ & 5.1 & 7.1 \nl
0234+285 & CTD 20 & 1.213 & $16\pm8$ & 6.6 & 13 \nl
1156+295 & 4C 29.45 & 0.729 & $41\pm2$ & $>2.1$ & $>4.9$ \nl
1253$-$055 & 3C 279 & 0.538 & $3.7\pm0.5$ & 13.2 & 14.0 \nl
1641+399 & 3C 345 & 0.595 & $9.6\pm0.2$ & 1.4 & 4.1 \nl
2223$-$052 & 3C 446 & 1.404 & $0.0\pm3.6$ & 16.6 & 16.0 \nl
2230+114 & CTA 102 & 1.037 & $0\pm24$ & $>0.9$ & $>1.5$ \nl
2251+158 & 3C 454.4 & 0.859 & $3.2\pm0.6$ & $>5.2$ & $>4.6$ \nl
\underline{CDLPQ}\\
0016+731 &... & 1.781 & $16\pm4$ & 5.2 & 7.9 \nl
0153+744 & ...& 2.340 & $1.9\pm3.5$ & $>1.3$ & $>1.8$ \nl
0333+321 & NRAO 140 & 1.258 & $8.3\pm0.5$ & 27 & 13.0 \nl
0430+052 & 3C 120 & 0.033 & $4.7\pm0.5$ & $>11$ & $>4.1$ \nl
0552+398 & DA 193 & 2.365 & $3.6\pm1.8$ & 1.1 & 2.2 \nl
0711+356 & OI 318 &1.620 & $0.0\pm1.7$ & 20 & 6.4 \nl
0836+710 & 4C 71.07 &2.170 & $16\pm3$ & 7.7 & 6.7 \nl
0923+392 & 4C 39.25 &0.699 & $6.2\pm0.3$ & 6.5 & 8.9 \nl
1226+023 & 3C 273 & 0.158 & $8.8\pm0.2$ & 7.9 & 4.6 \nl
1928+738 & 4C 73.18 & 0.302 & $6.8\pm0.3$ & 3.2 & 3.4 \nl
\underline{CDQ-NPI}\\
0615+820 &... & 0.710 & $1.7\pm1.7$ & $0.7$ & $1.4$ \nl
1039+811 &... & 1.260 & $<4$ & $16.6$ & $12.2$ \nl
1150+812 &... & 1.250 & $6\pm3$ & $8.2$ & $9.6$ \tablebreak
\underline{LDQ}\\
0850+581 & 4C 58.17 & 1.322 & $7\pm1$ & 2.2 & 2.5 \nl
0906+430 & 3C 216 & 0.670 & $6\pm1$ & 61 & 33 \nl
1040+123 & 3C 245 & 1.029 & $5.2\pm2.5$ & 0.5 & 1.4 \nl
1222+216 & 4C 21.35 & 0.435 & $2.1\pm0.9$ & 1.2 & 1.0 \nl
1618+177 & 3C 334 & 0.555 & $2.8\pm0.9$ & $>0.2$ & $>0.3$ \nl
1721+343 & 4C 34.47 & 0.206 & $2.9\pm0.3$ & 0.1 & 0.2 \nl
1830+285 & 4C 28.45 & 0.594 & $3.9\pm0.8$ & 0.4 & 0.4 \nl
\underline{RG}\\
0108+388 & OC 314 & 0.669 & $0.90\pm0.25$ & 0.2 & 0.7 \nl
0316+413 & 3C 84 & 0.018 & $0.6\pm0.1$ & 1.2 & 1.2 \nl
0710+439 &... & 0.518 & $0.1\pm0.2$ & 0.2 & 0.4 \nl
1228+127 & M 87 & 0.004 & $0.2\pm0.1$ & 0.8 & 0.8 \nl
1637+826 & NGC 6251 & 0.023 & $0.1\pm0.1$ & $>1.3$ & $>1.0$ \nl
2021+614 & OW 637 & 0.227 & $0.3\pm0.3$ & 0.9 & 1.1 \nl
2352+495 & OZ 488 & 0.237 & $<0.4$ & 0.3 & 0.5 \nl
\enddata
\end{deluxetable}

\begin{deluxetable}{ccccccc}
\scriptsize
\tablecaption{Solutions for $\Gamma$ and $\phi$
\label{tb:out}}
\tablewidth{0pt}
\tablehead{
\colhead{Source} & \colhead{Name} &  \colhead{$z$} &
\colhead{$\Gamma_{eq}$} &
\colhead{$\phi_{eq}$(deg)} & \colhead{$\Gamma_{IC}$} &
\colhead{$\phi_{IC}$(deg)} \\
\colhead{(1)} & \colhead{(2)} & \colhead{(3)} & \colhead{(4)} &
\colhead{(5)} & \colhead{(6)} & \colhead{(7)}
}
\startdata
\underline{BL Lacs}\\
0454+844 &... & 0.112 & $1.4\pm0.2$ &
$48\pm97$ & $1.6\pm0.8$ & $18\pm35$ \nl 
0851+202 & OJ 287 & 0.306 & $6\pm8$ &
$3\pm9$ & $5\pm2$ & $7\pm11$ \nl
1101+384 & Mkn 421 & 0.031 & $3\pm1$ &
$26\pm34$ & $5\pm4$ & $39\pm7$ \nl
1308+326 &... & 0.996 & $42\pm72$ &
$6\pm3$ & $33\pm40$ & $6\pm3$ \nl
1749+701 &... & 0.770 & $83\pm133$ &
$12\pm1$ & $56\pm56$ & $12\pm1$ \nl
1803+784 &... & 0.684 & $3\pm4$ &
$0.6\pm4.5$ & $3\pm3$ & $0.4\pm2.6$ \nl
2007+776 &... & 0.342 & $3.6\pm1.6$ &
$21\pm26$ & $3.5\pm0.7$ & $16\pm17$ \nl
2200+420 & BL Lac & 0.069 & $4.7\pm0.4$ &
$12\pm20$ & $5.0\pm1.6$ & $16\pm11$ \nl
\underline{CDHPQ}\\
0212+735 &... & 2.370 & $9\pm9$ &
$10\pm10$ & $8\pm5$ & $8\pm7$ \nl
0234+285 & CTD 20 & 1.213 & $23\pm32$ &
$6\pm4$ & $16\pm10$ & $4\pm3$ \nl
1156+295 & 4C 29.45 & 0.729 & $<405$ &
$<2.8$ & $<175$ & $<2.8$ \nl
1253$-$055 & 3C 279 & 0.538 & $7\pm10$ &
$2\pm7$ & $7.5\pm6.5$ & $2\pm4$ \nl
1641+399 & 3C 345 & 0.595 & $34\pm52$ &
$12\pm1$ & $13\pm9$ & $10\pm3$ \nl
2223$-$052 & 3C 446 & 1.404 & $8\pm13$ &
$0.0\pm1.5$ & $8\pm8$ & $0.0\pm1.6$ \nl
2230+114 & CTA 102 & 1.037 & $>1.0$ &
$0\pm180$ & $>1.1$ & $0\pm180$ \nl
2251+158 & 3C 454.4 & 0.859 & $>3.7$ &
$<10$ & $>3.5$ & $<12$ \nl
\underline{CDLPQ}\\
0016+731 &... & 1.781 & $27\pm36$ &
$7\pm2$ & $20\pm14$ & $6\pm2$ \nl
0153+744 & ...& 2.340 & $>2.1$ &
$<42$ & $>2.1$ & $<33$ \nl
0333+321 & NRAO 140 & 1.258 & $15\pm20$ &
$1\pm4$ & $9\pm4$ & $4\pm6$ \nl
0430+052 & 3C 120 & 0.033 & $>6$ &
$<4$ & $>4.5$ & $<14$ \nl
0552+398 & DA 193 & 2.365 & $7\pm11$ &
$29\pm14$ & $4.2\pm3.5$ & $23\pm13$ \nl
0711+356 & OI 318 &1.620 & $10\pm16$ &
$0\pm1$ & $3\pm3$ & $0\pm5$ \nl
0836+710 & 4C 71.07 &2.170 & $19\pm20$ &
$6\pm4$ & $21\pm16$ & $6\pm2$ \nl
0923+392 & 4C 39.25 &0.699 & $6.3\pm0.5$ &
$9\pm15$ & $6.6\pm2.3$ & $6\pm8$ \nl
1226+023 & 3C 273 & 0.158 & $8.9\pm1.5$ &
$7\pm10$ & $11\pm6$ & $10\pm4$ \nl
1928+738 & 4C 73.18 & 0.302 & $8.9\pm9.1$ &
$14\pm8$ & $8.6\pm5.2$ & $14\pm5$ \nl
\underline{CDQ-NPI}\\
0615+820 &... & 0.710 & $3\pm6$ &
$54\pm42$ & $2\pm2$ & $41\pm31$ \nl
1039+811 &... & 1.260 & $<8.8$ &
$<1.5$ & $<6.7$ & $<2.7$ \nl
1150+812 &... & 1.250 & $6\pm4$ &
$7\pm14$ & $7\pm3$ & $5\pm8$ \tablebreak
\underline{LDQ}\\
0850+581 & 4C 58.17 & 1.322 & $12\pm16$ &
$15\pm5$ & $11\pm10$ & $15\pm4$ \nl
0906+430 & 3C 216 & 0.670 & $31\pm49$ &
$0.2\pm0.6$ & $17\pm16$ & $0.6\pm1.2$ \nl
1040+123 & 3C 245 & 1.029 & $28\pm51$ &
$22\pm10$ & $11\pm13$ & $21\pm9$ \nl
1222+216 & 4C 21.35 & 0.435 & $2.8\pm2.9$ &
$41\pm30$ & $3.2\pm2.9$ & $44\pm18$ \nl
1618+177 & 3C 334 & 0.555 & $>3$ &
$<39$ & $>3$ & $<39$ \nl
1721+343 & 4C 34.47 & 0.206 & $46\pm74$ &
$38\pm4$ & $32\pm32$ & $38\pm4$ \nl
1830+285 & 4C 28.45 & 0.594 & $20\pm32$ &
$29\pm6$ & $21\pm23$ & $29\pm6$ \nl
\underline{RG}\\
0108+388 & OC 314 & 0.669 & $4\pm6$ &
$94\pm17$ & $1.7\pm1.0$ & $81\pm32$ \nl
0316+413 & 3C 84 & 0.018 & $1.15\pm0.07$ &
$60\pm160$ & $1.15\pm0.08$ & $56\pm99$ \nl
0710+439 &... & 0.518 & $2.6\pm3.9$ &
$170\pm25$ & $1.5\pm1.1$ & $170\pm30$ \nl 
1228+127 & M 87 & 0.004 & $1.0\pm0.4$ &
$120\pm180$ & $1.1\pm0.3$ & $130\pm90$ \nl
1637+826 & NGC 6251 & 0.023 & $>1.1$ &
$<15$ & $>1.0$ & $<88$ \nl
2021+614 & OW 637 & 0.227 &  $1.1\pm0.3$ &
$110\pm180$ & $1.04\pm0.09$ & $60\pm180$ \nl
2352+495 & OZ 488 & 0.237 &  $<1.8$ &
$>133$ & $<1.25$ & $>124$ 
\enddata
\end{deluxetable}

\begin{deluxetable}{ccccc}
\tablecaption{Median Values of $\Gamma$ and $\phi$ \label{tb:med}}
\tablewidth{0pt}
\tablehead{
\colhead{Class} & 
\colhead{Med. $\Gamma_{eq}$} & 
\colhead{Med. $\phi_{eq}$(deg)} & \colhead{Med. $\Gamma_{IC}$} &
\colhead{Med. $\phi_{IC}$(deg)}
}
\startdata
BL Lacs  & $4.2\pm0.7$ & $12\pm4$ & $5.0\pm0.8$ & $14\pm2$ \nl
CDHPQ    & $9\pm1$ & $4\pm2$ & $8\pm1$ & $3.0\pm1.5$ \nl
CDLPQ    & $9.5\pm1.2$ & $7\pm2$ & $8.8\pm1.7$ & $6\pm1$ \nl
CDQ-NPI  & $4.5\pm1.5$ & $30\pm23$ & $4.5\pm2.5$ & $23\pm18$ \nl
LDQ      & $24\pm5$ & $26\pm6$ & $14\pm3$ & $25\pm6$ \nl
RG       & $1.15\pm0.08$ & $110\pm9$ & $1.15\pm0.06$ & $81\pm14$ \nl
\enddata
\end{deluxetable}

\clearpage

\figcaption{(a) The equipartition Doppler factor, $\delta_{eq}$, vs. $(1+z)$;
(b) the inverse Compton Doppler factor, $\delta_{IC}$, vs. 
$(1+z)$ for the 100 sources in this study.
Symbols: solid circles are BL Lacs, solid diamonds are CDHPQs,
solid squares are CDLPQs, solid triangles are CDQ-NPIs,
open diamonds are LDQs, and open squares are RGs.
\label{fig:deleq}}

\figcaption{The distribution of equipartition Doppler factors,
$\delta_{eq}$ (dotted line), and inverse Compton Doppler factors,
$\delta_{IC}$ (dashed line), for all sources in this study.
\label{fig:dophist}}

\figcaption{The distribution of equipartition Doppler factors,
$\delta_{eq}$ (dotted line), and inverse Compton Doppler factors,
$\delta_{IC}$ (dashed line), for each class of AGN: (a) BL Lacs,
(b) CDHPQs, (c) CDLPQs, (d) CDQ-NPIs, (e) LDQs, (f) RGs.
\label{fig:dophistcl}}

\figcaption{The distribution of observed brightness
temperatures, $T_{Bo}$ (solid line),
and intrinsic brightness temperatures, $T_{Bi}(eq)$ (dotted line) 
\& $T_{Bi}(IC)$
(dashed line), 
for all sources in this study.
\label{fig:temphist}}

\figcaption{The distribution of observed brightness
temperatures, $T_{Bo}$ (solid line),
and intrinsic brightness temperatures, $T_{Bi}(eq)$ (dotted line) \&
$T_{Bi}(IC)$
(dashed line),
for each class of AGN: (a) BL Lacs,
(b) CDHPQs, (c) CDLPQs, (d) CDQ-NPIs, (e) LDQs, (f) RGs.
\label{fig:temphistcl}}

\figcaption{The uncorrected luminosity density, $L_{\nu_{un}}$, vs. $(1+z)$
for all 100 sources.
Symbols: solid circles are BL Lacs, solid diamonds are CDHPQs,
solid squares are CDLPQs, solid triangles are CDQ-NPIs,
open diamonds are LDQs, and open squares are RGs.
\label{fig:oblumz}}

\figcaption{The uncorrected luminosity , $L_{un}$, vs. $(1+z)$
for all 100 sources.
Symbols: solid circles are BL Lacs, solid diamonds are CDHPQs,
solid squares are CDLPQs, solid triangles are CDQ-NPIs,
open diamonds are LDQs, and open squares are RGs.
\label{fig:oblumintz}}

\figcaption{The intrinsic luminosity density $L_{\nu i}(eq)$
 vs. the rest frame frequency $\nu_{i}(eq)$.  The result of
a power law fit is $L_{\nu i}(eq)\propto\nu_{i}(eq)^{2.3\pm0.1}$
with a reduced $\chi^2$ equal to 4.6 and a correlation coefficient $r=0.91$.
Symbols: solid circles are BL Lacs, solid diamonds are CDHPQs,
solid squares are CDLPQs, solid triangles are CDQ-NPIs,
open diamonds are LDQs, and open squares are RGs.
\label{fig:lumeq}}

\figcaption{The intrinsic luminosity $L_{i}(eq)$ 
vs. the rest frame frequency $\nu_{i}(eq)$.
The result of
a power law fit is $L_{i}(eq)\propto\nu_{i}(eq)^{3.3\pm0.1}$
with a reduced $\chi^2$ equal to 4.0 and a correlation coefficient $r=0.96$.
Symbols: solid circles are BL Lacs, solid diamonds are CDHPQs,
solid squares are CDLPQs, solid triangles are CDQ-NPIs,
open diamonds are LDQs, and open squares are RGs.
\label{fig:luminteq}} 

\figcaption{The intrinsic luminosity density $L_{\nu i}(eq)$
vs. $(1+z)$.  
The result of
a power law fit is $L_{\nu i}(eq)\propto (1+z)^{2.5\pm1.6}$
with a reduced $\chi^2$ equal to 1.3 and a correlation coefficient $r=0.16$.
Symbols: solid circles are BL Lacs, solid diamonds are CDHPQs,
solid squares are CDLPQs, solid triangles are CDQ-NPIs,
open diamonds are LDQs, and open squares are RGs.
\label{fig:lumz}}

\figcaption{The intrinsic luminosity $L_{i}(eq)$ 
vs. $(1+z)$.
The result of
a power law fit is $L_{\nu i}(eq)\propto (1+z)^{0.9\pm2.2}$
with a reduced $\chi^2$ equal to 1.4 and a correlation coefficient $r=0.04$.
Symbols: solid circles are BL Lacs, solid diamonds are CDHPQs,
solid squares are CDLPQs, solid triangles are CDQ-NPIs,
open diamonds are LDQs, and open squares are RGs.
\label{fig:lumintz}}

\figcaption{The apparent speed in the plane of the sky, $\beta_{app}$, vs. 
redshift, $z$ for the overlap of the GPCM93 and VC94 samples (upper bounds
are indicated by arrows). 
Symbols: solid circles are BL Lacs, solid diamonds are CDHPQs,
solid squares are CDLPQs, solid triangles are CDQ-NPIs,
open diamonds are LDQs, and open squares are RGs.
\label{fig:betapp}}

\figcaption{$\Gamma_{eq}$ vs. $\phi_{eq}$, estimates using the
equipartition Doppler factor for the overlap of the GPCM93 and VC94
samples (upper and lower bounds indicated by arrows): (a) with errors
shown, and (b) without errors shown.
Symbols: solid circles are BL Lacs, solid diamonds are CDHPQs,
solid squares are CDLPQs, solid triangles are CDQ-NPIs,
open diamonds are LDQs, and open squares are RGs.
\label{fig:gamtheeq}}

\figcaption{$\Gamma_{IC}$ vs. $\phi_{IC}$, estimates using the
inverse Compton Doppler factor for the overlap of the GPCM93  and VC94
samples (upper and lower bounds indicated by arrows): (a) with errors
shown, and (b) without errors shown.
\label{fig:gamtheic}}

\figcaption{(a) $\phi_{eq}$ and (b) $\Gamma_{eq}$ 
as functions of $(1+z)$.
Symbols: solid circles are BL Lacs, solid diamonds are CDHPQs,
solid squares are CDLPQs, solid triangles are CDQ-NPIs,
open diamonds are LDQs, and open squares are RGs.
\label{fig:zeq}}

\figcaption{(a) $\phi_{IC}$ and (b) $\Gamma_{IC}$ 
as functions of $(1+z)$.
Symbols: solid circles are BL Lacs, solid diamonds are CDHPQs,
solid squares are CDLPQs, solid triangles are CDQ-NPIs,
open diamonds are LDQs, and open squares are RGs.
\label{fig:zic}}

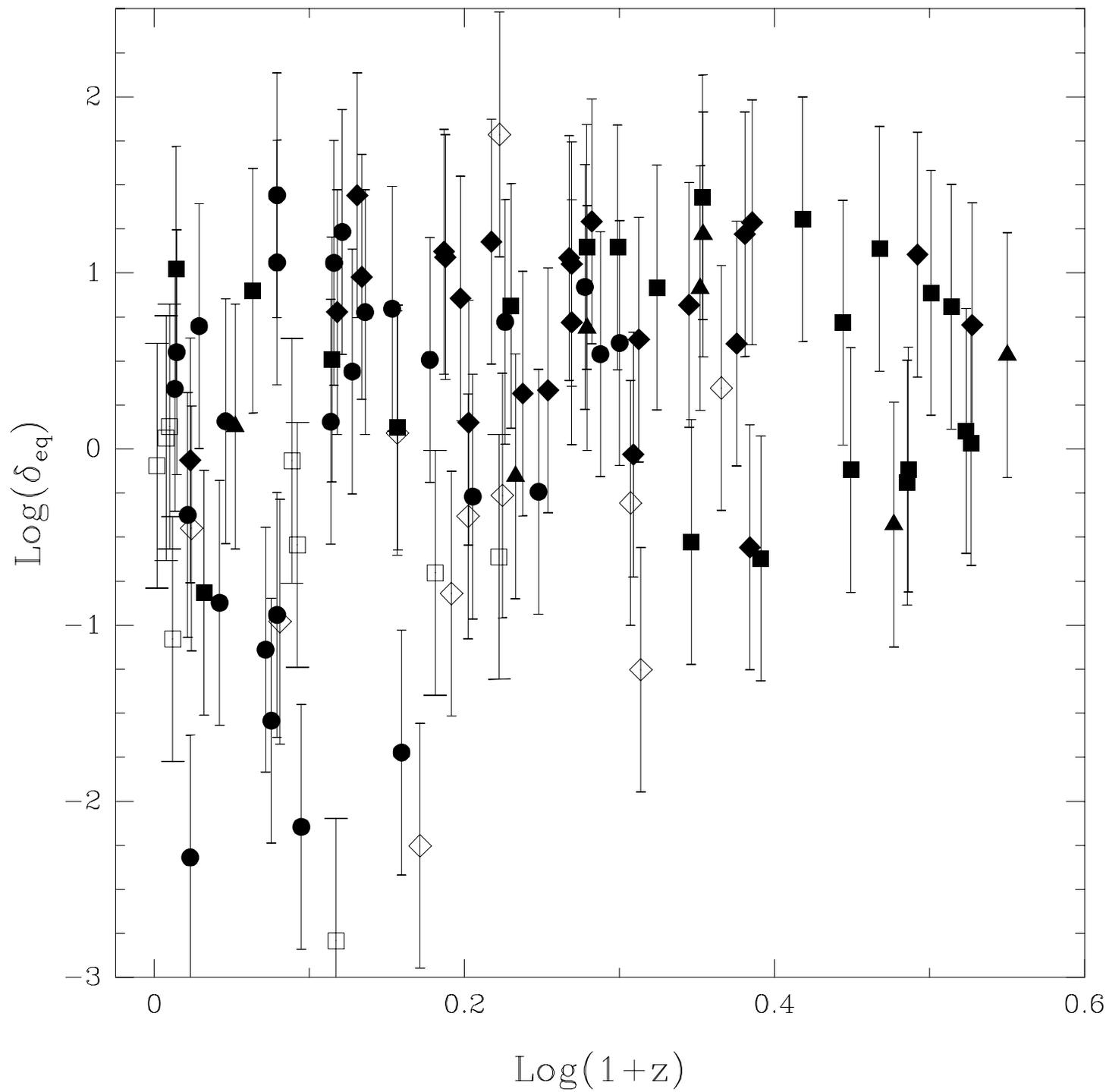
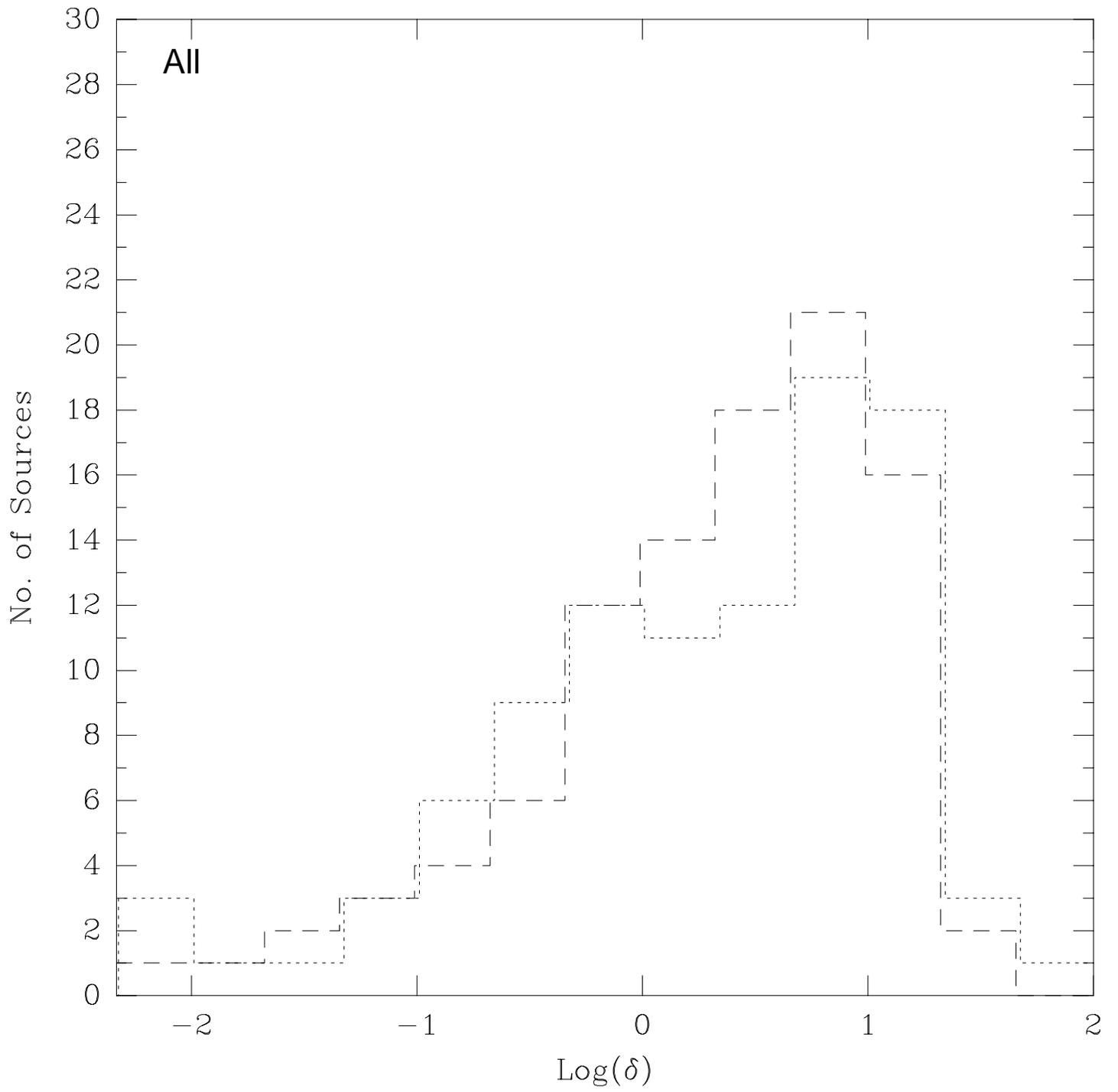
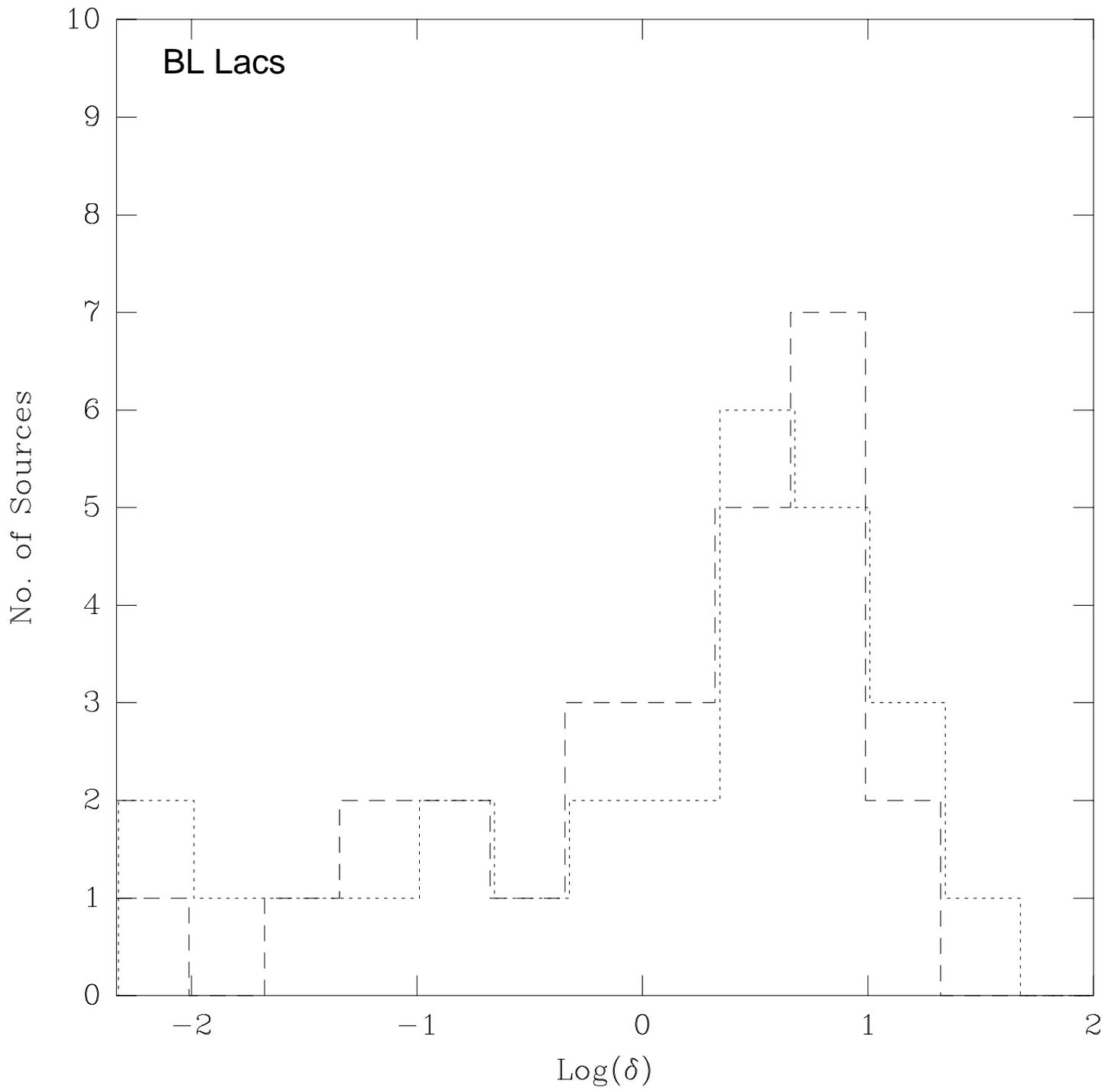
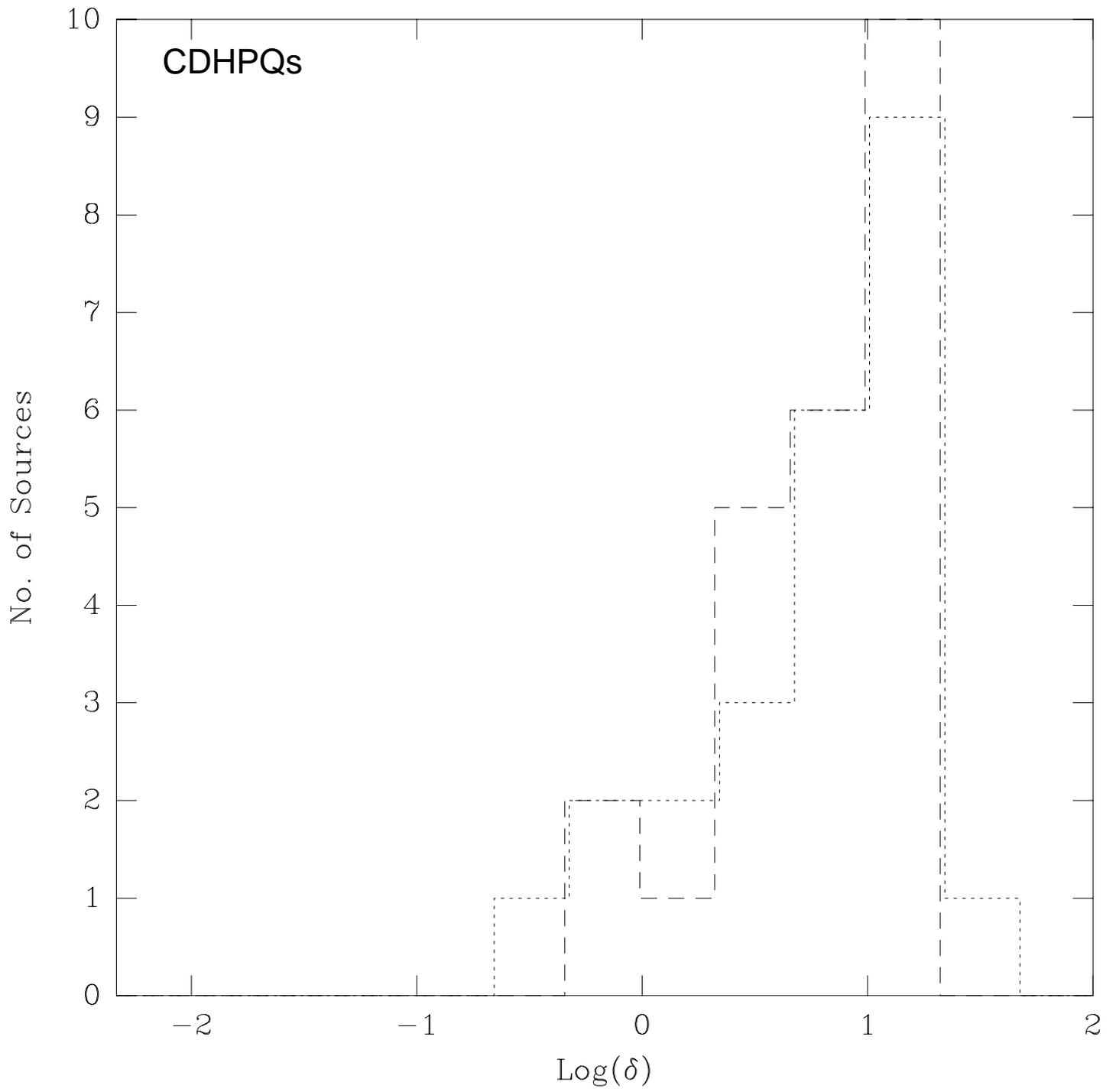
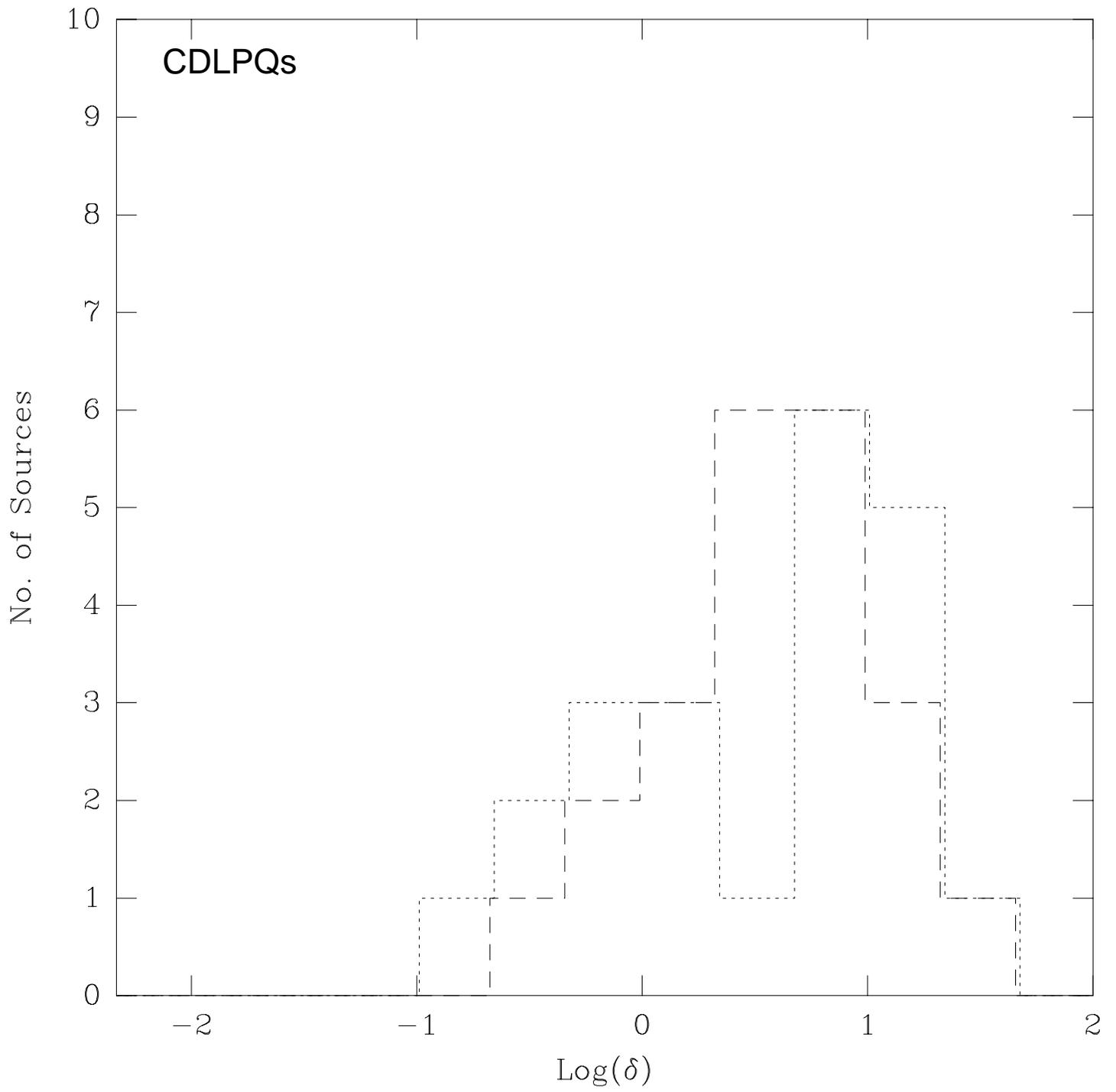
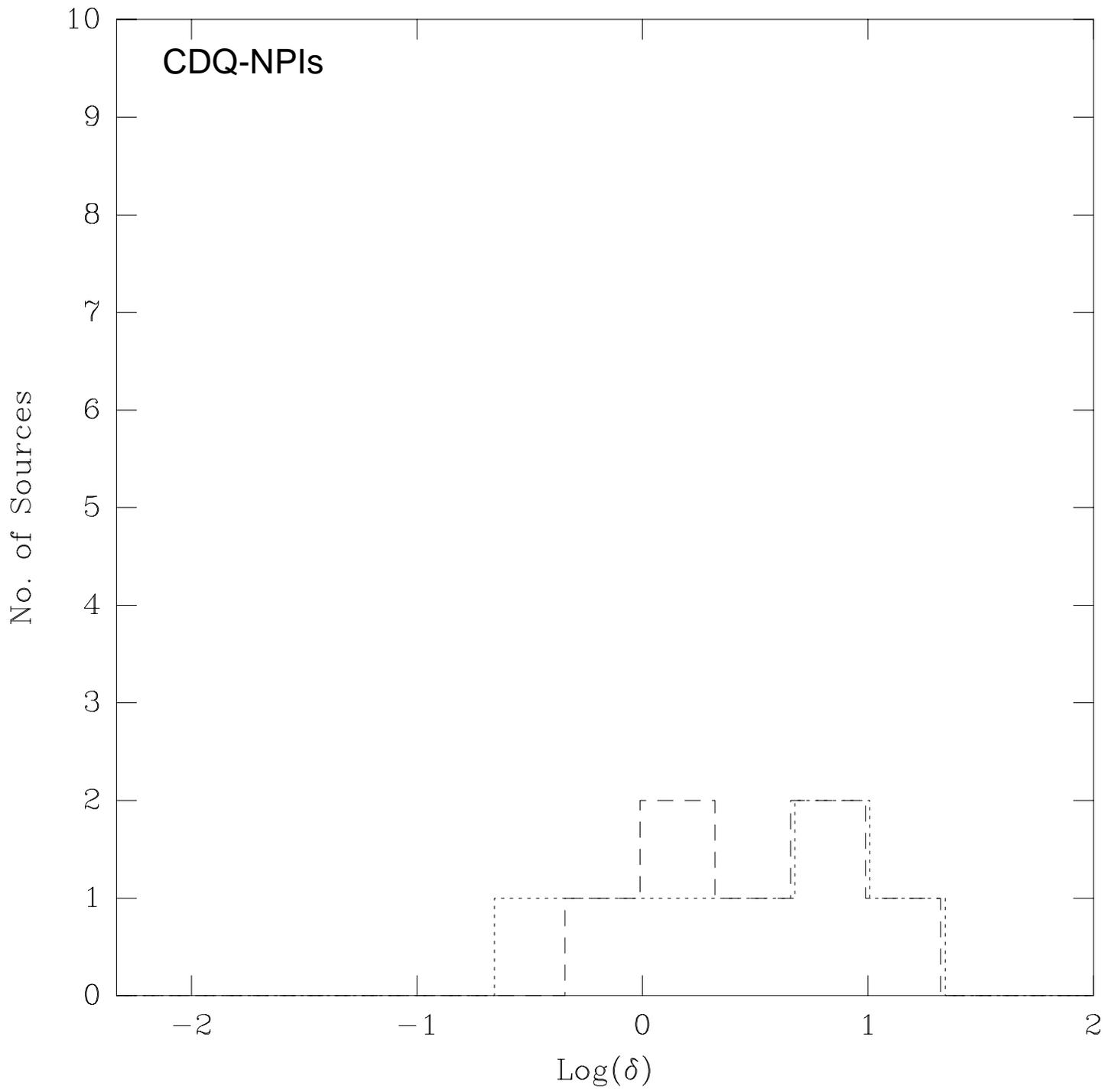
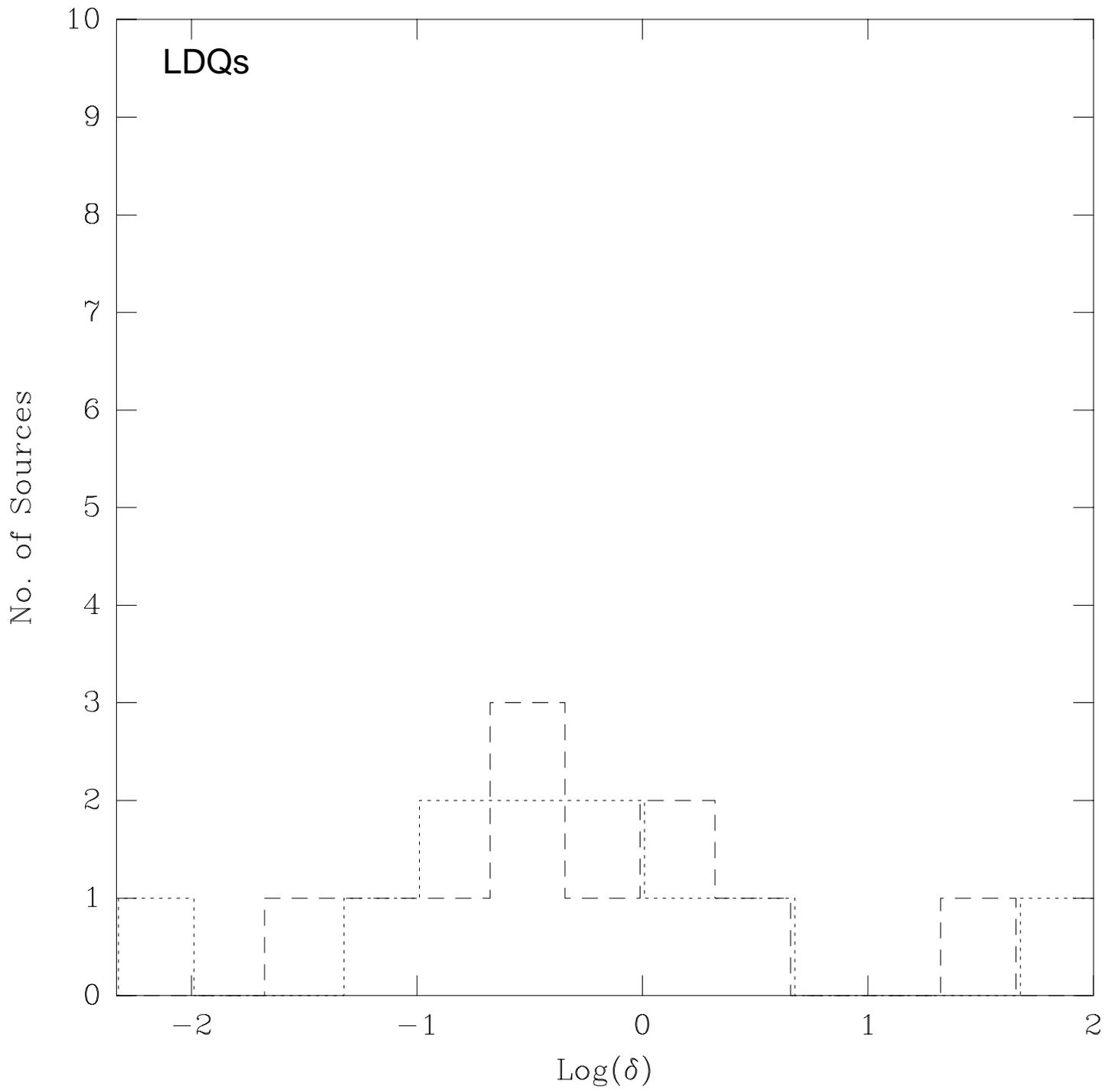
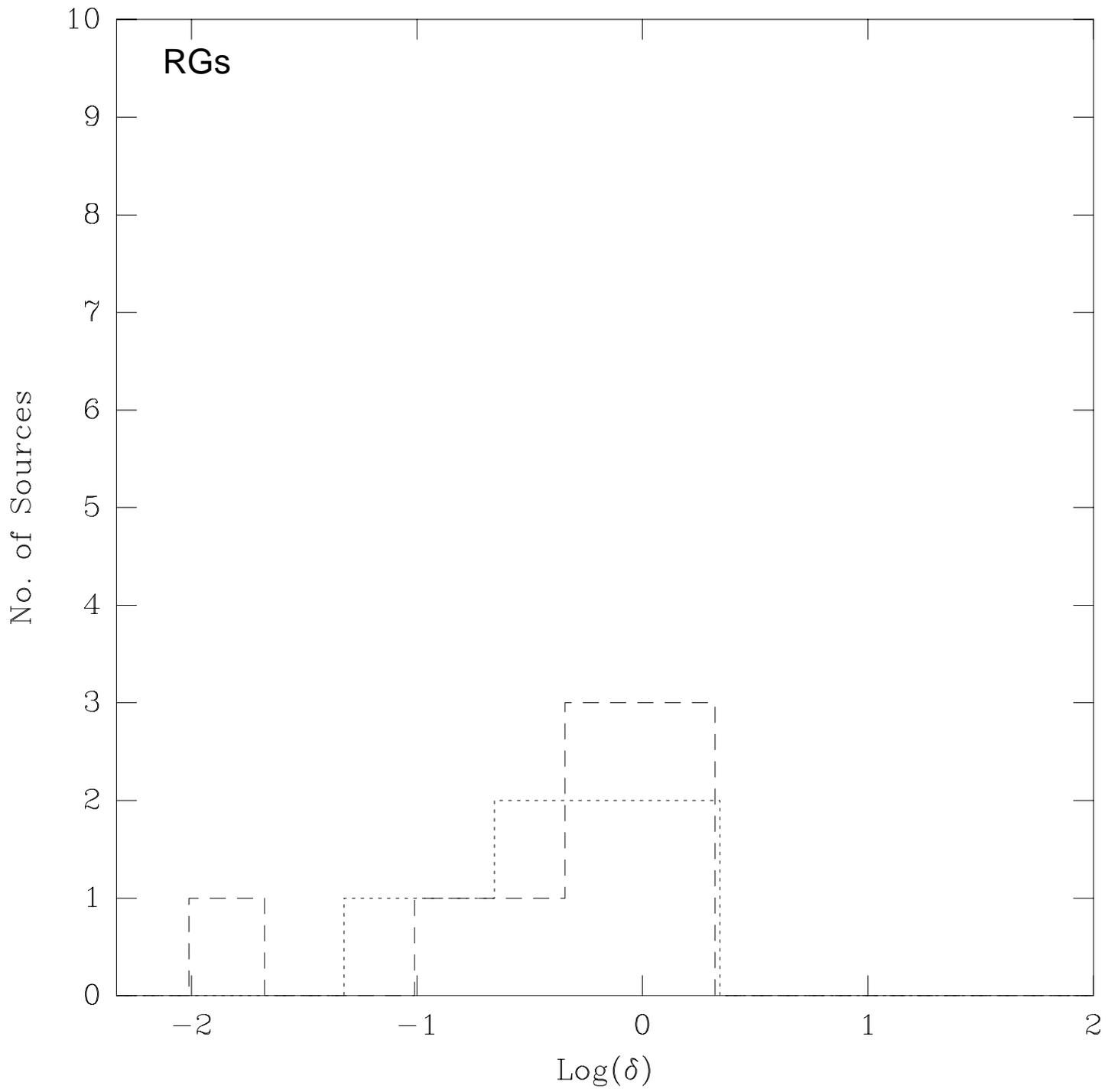
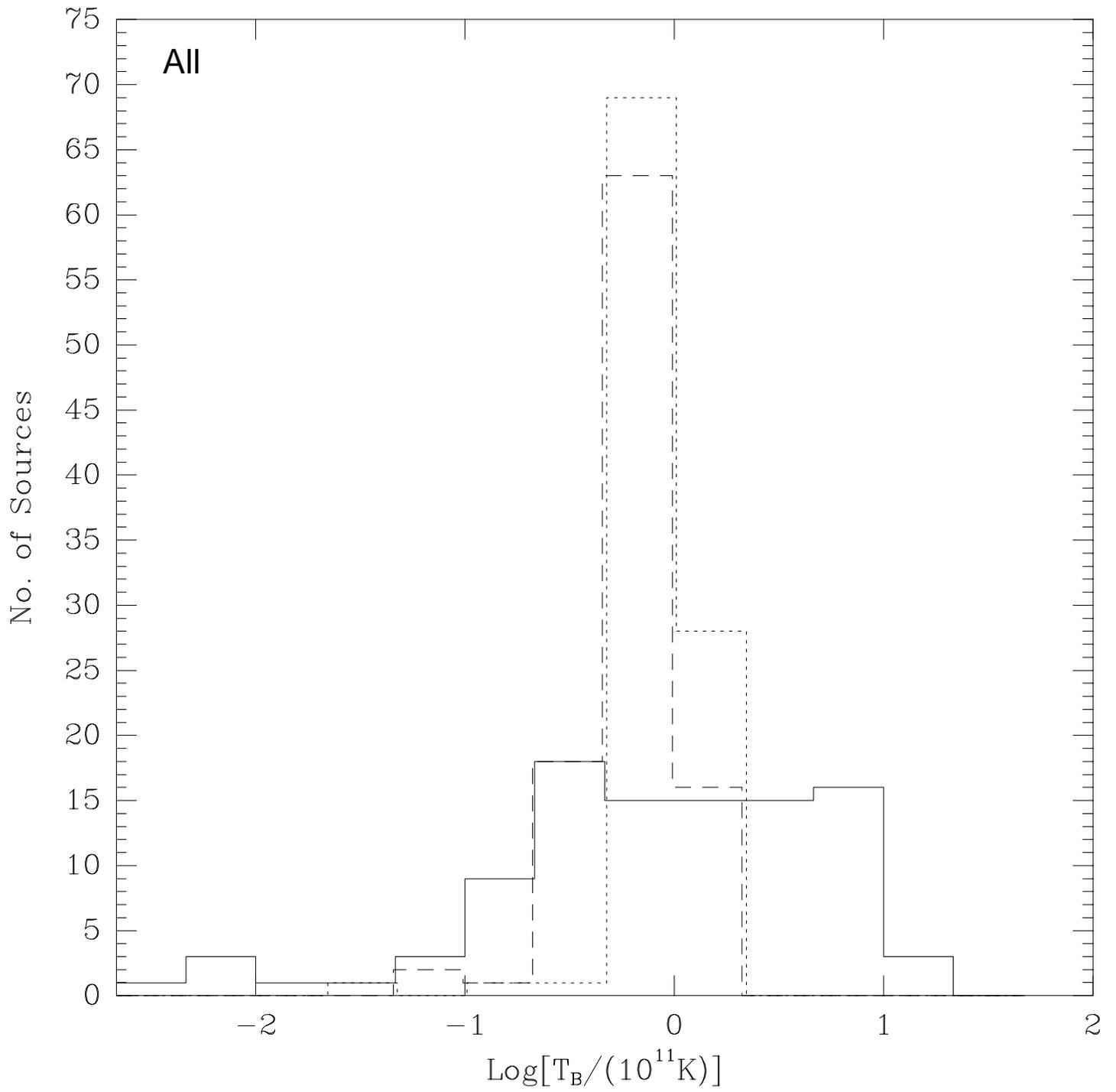
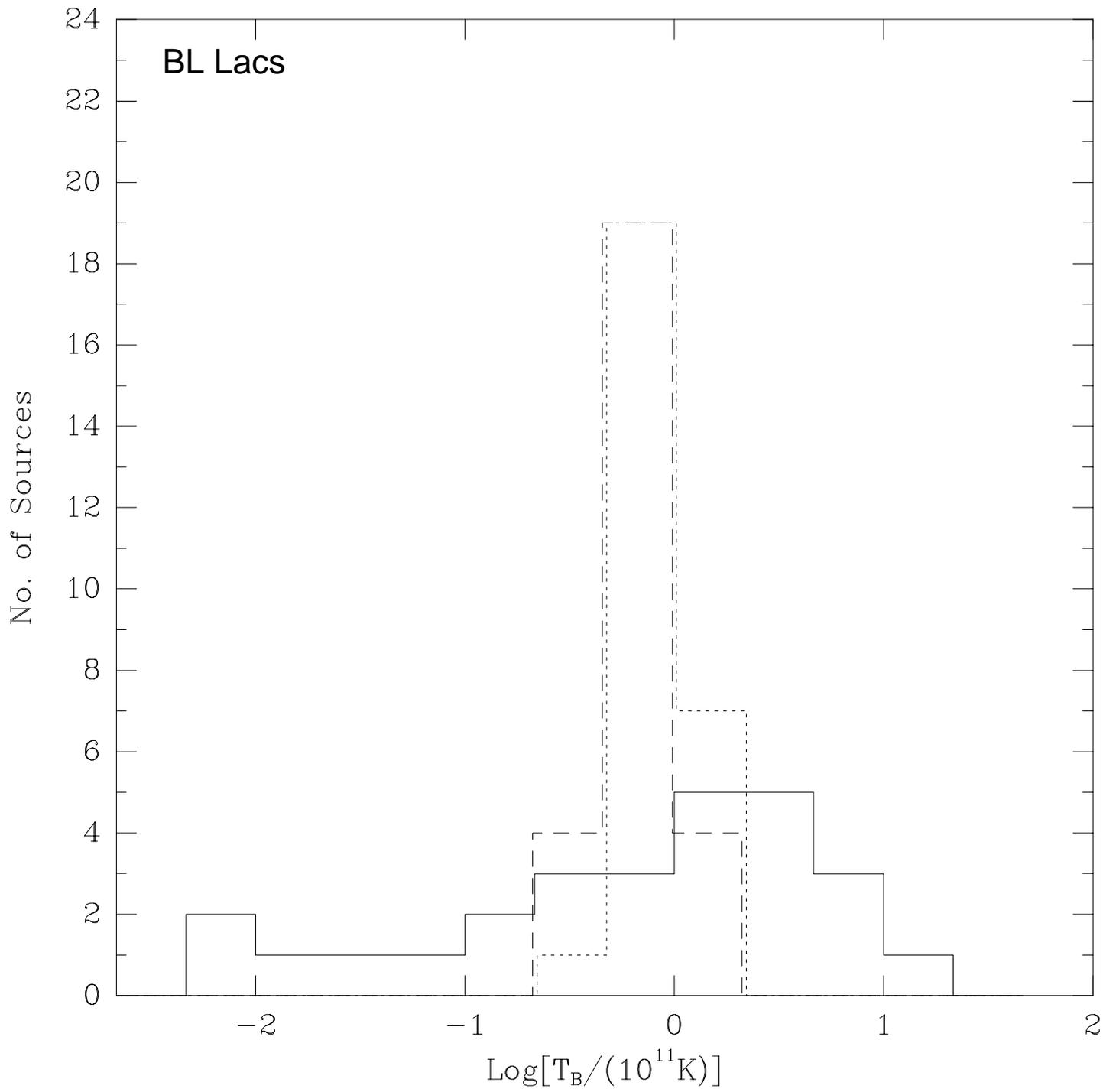
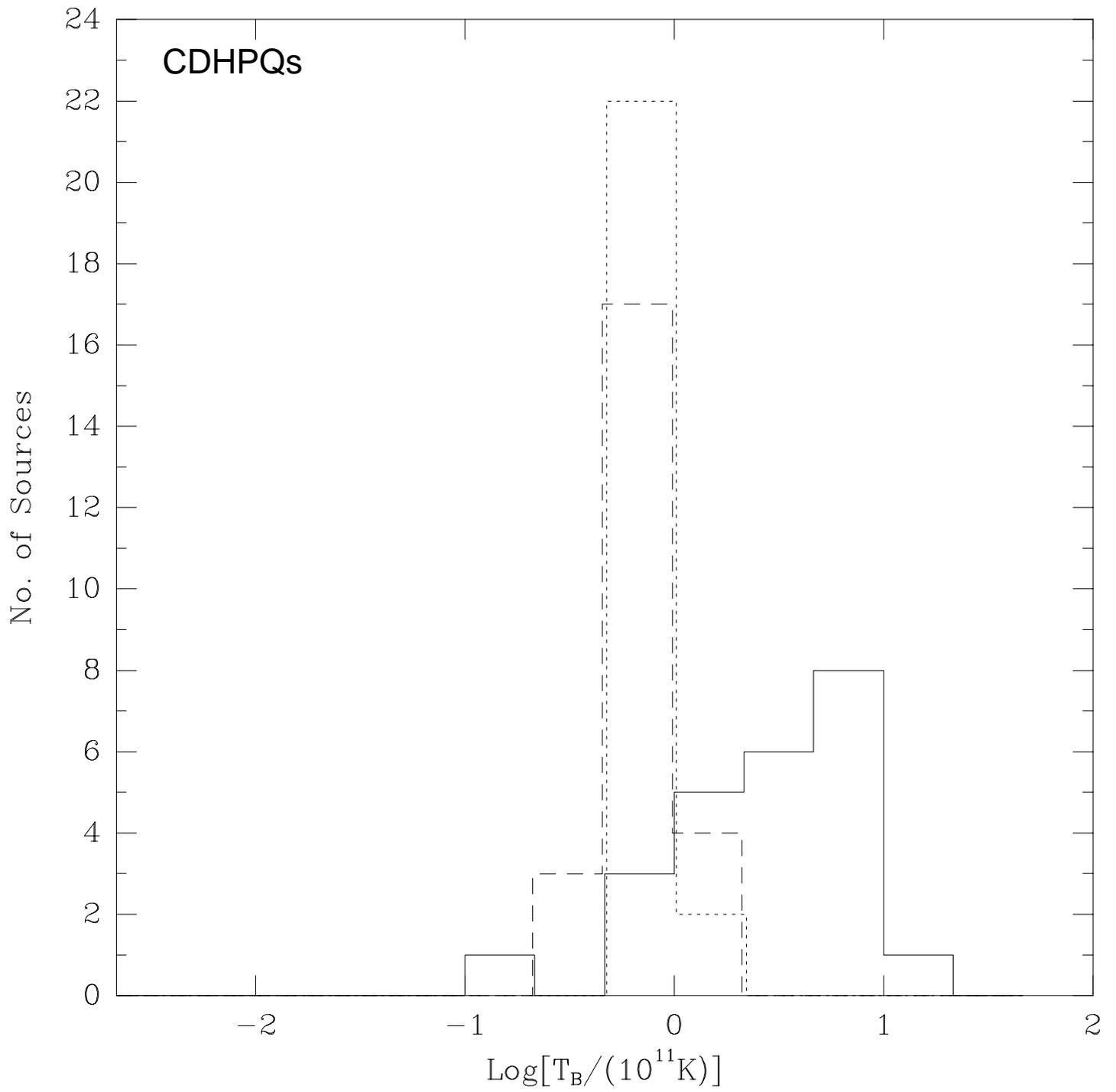
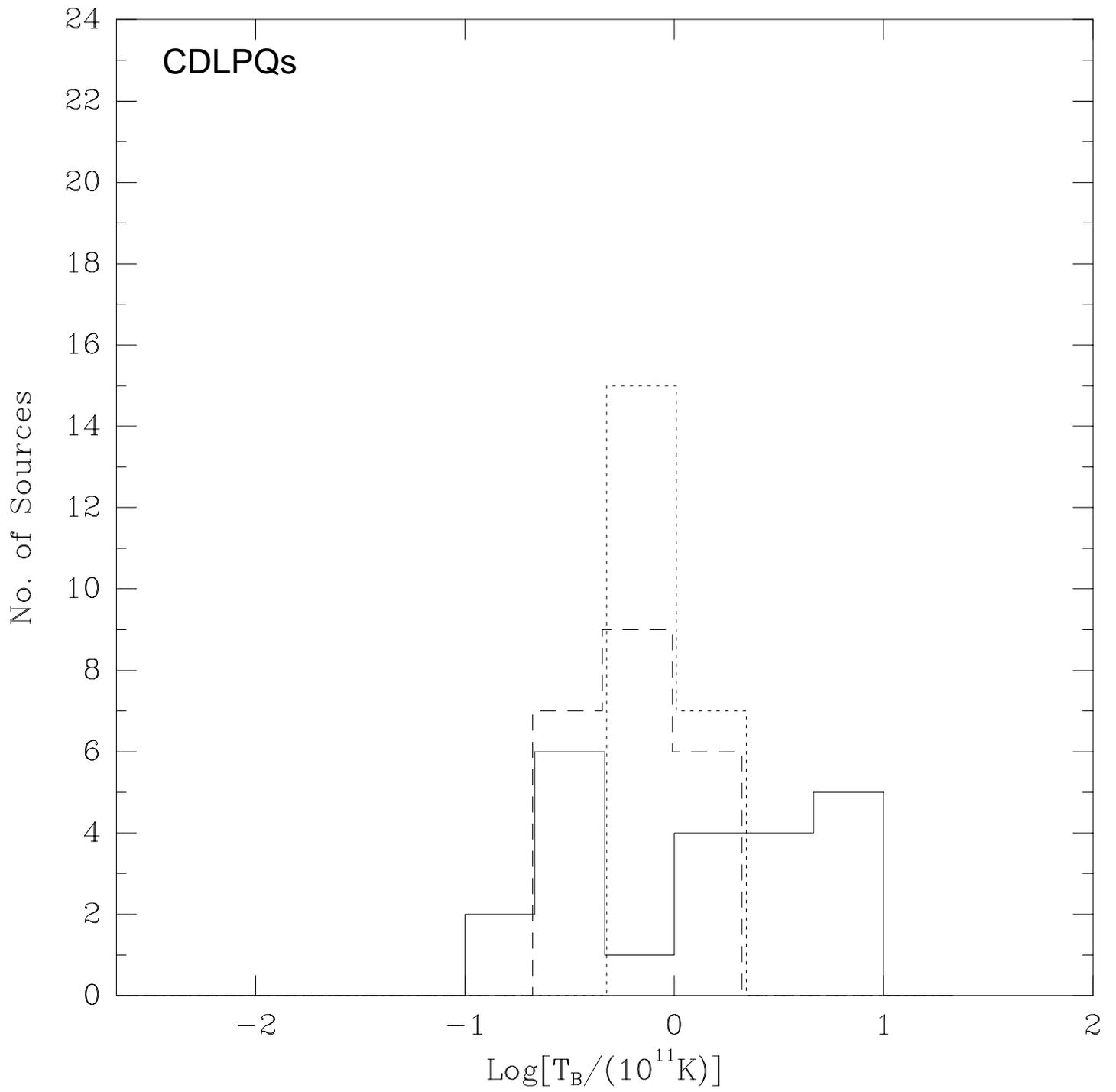
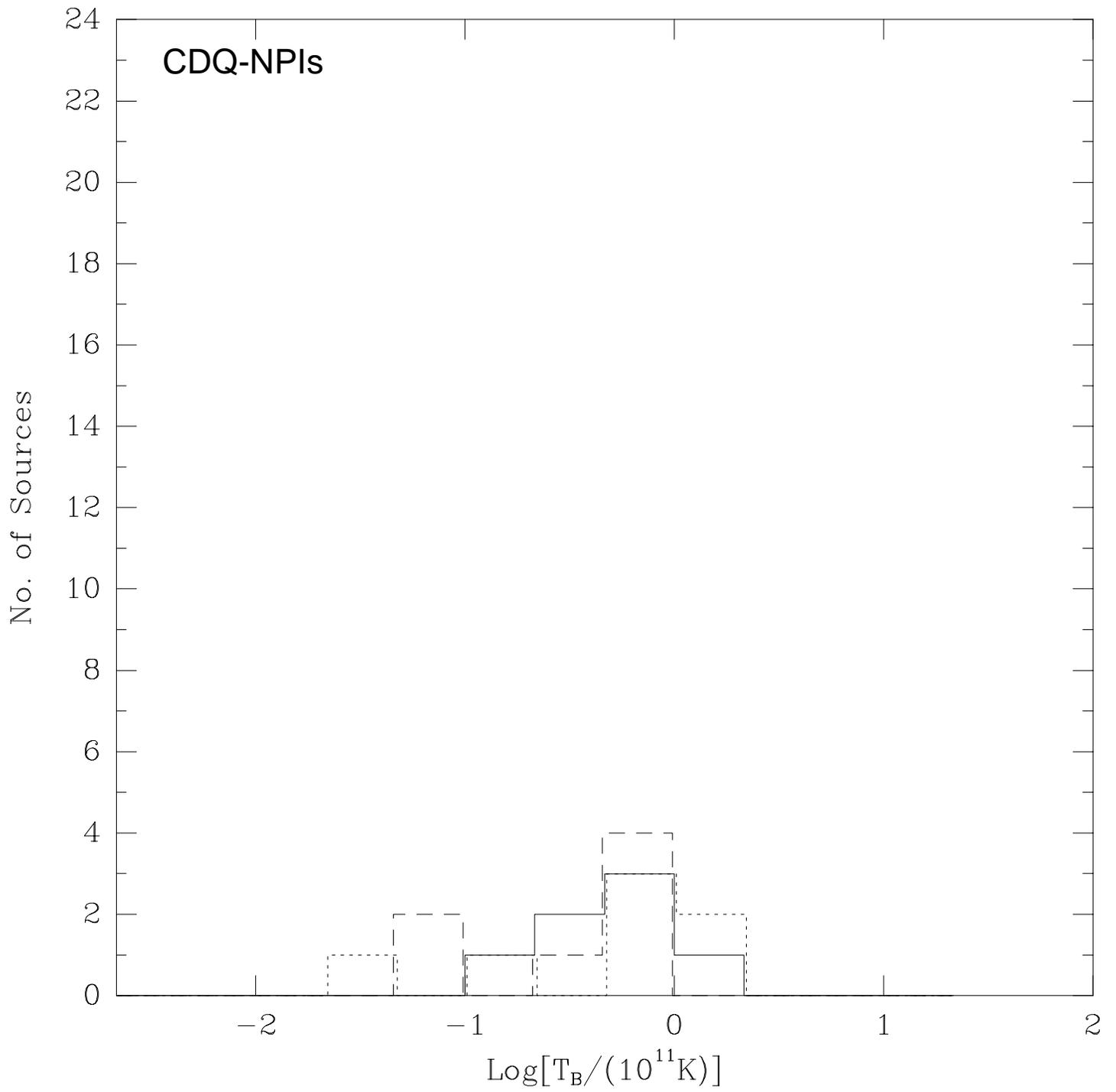
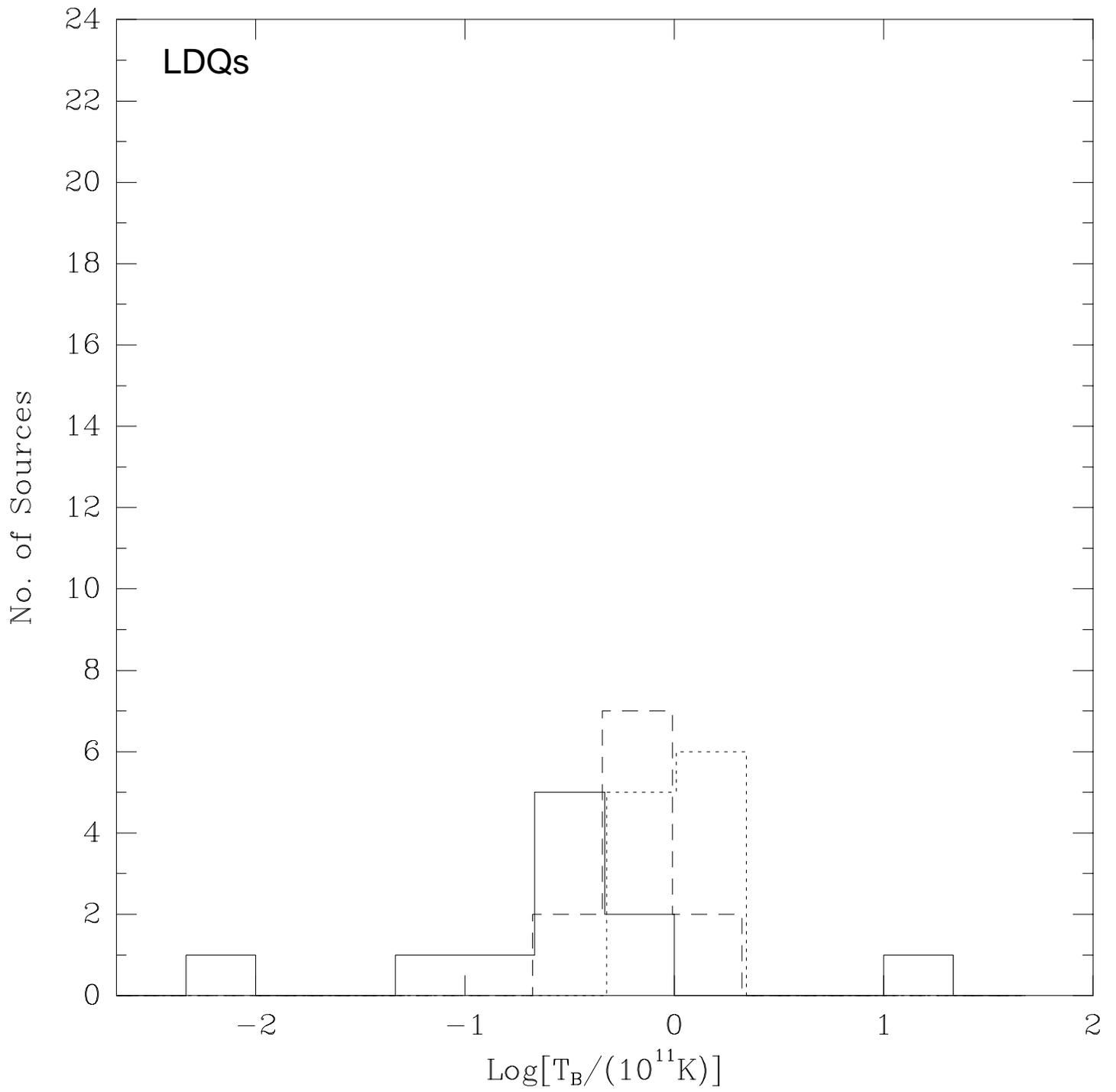
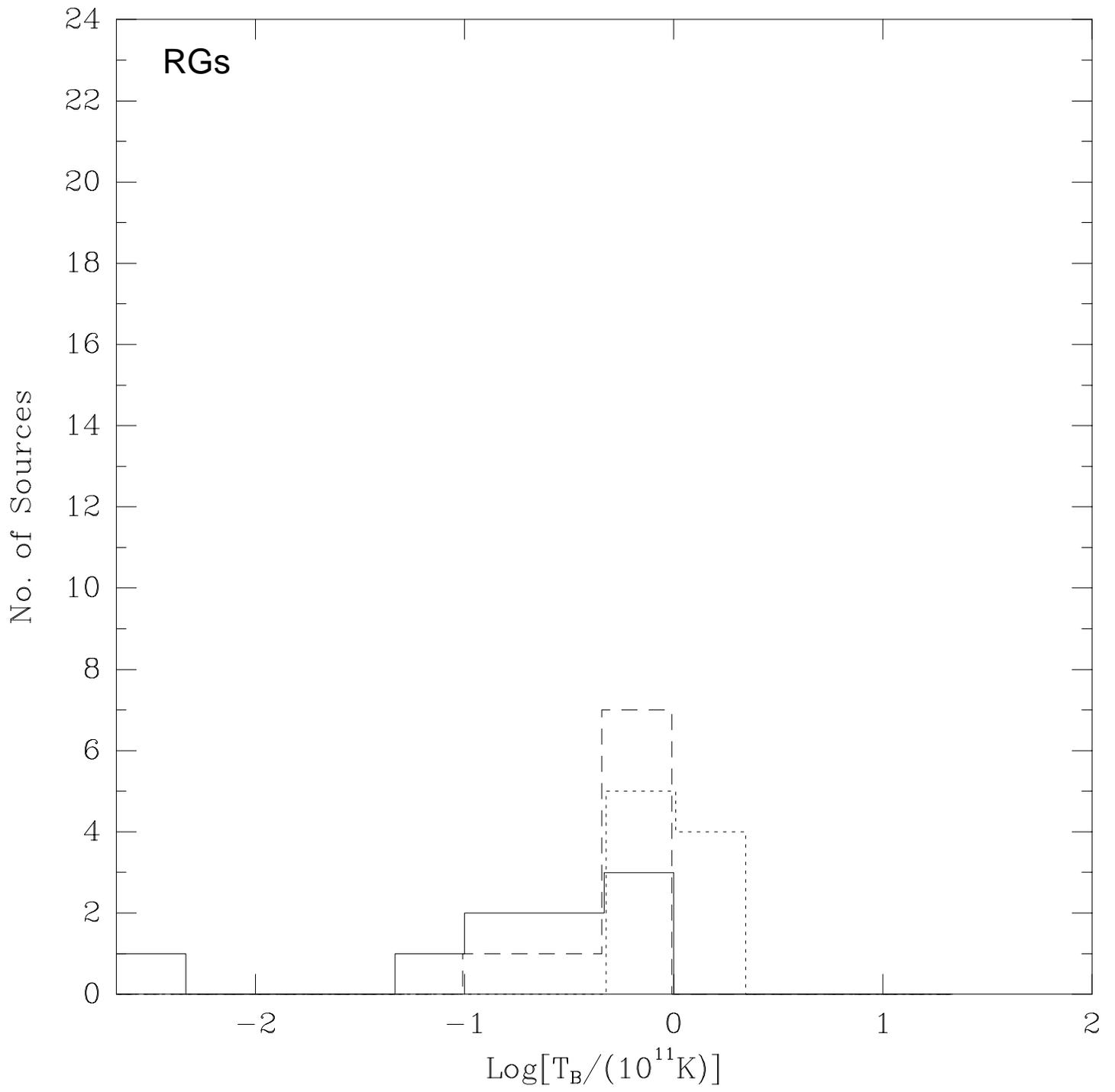
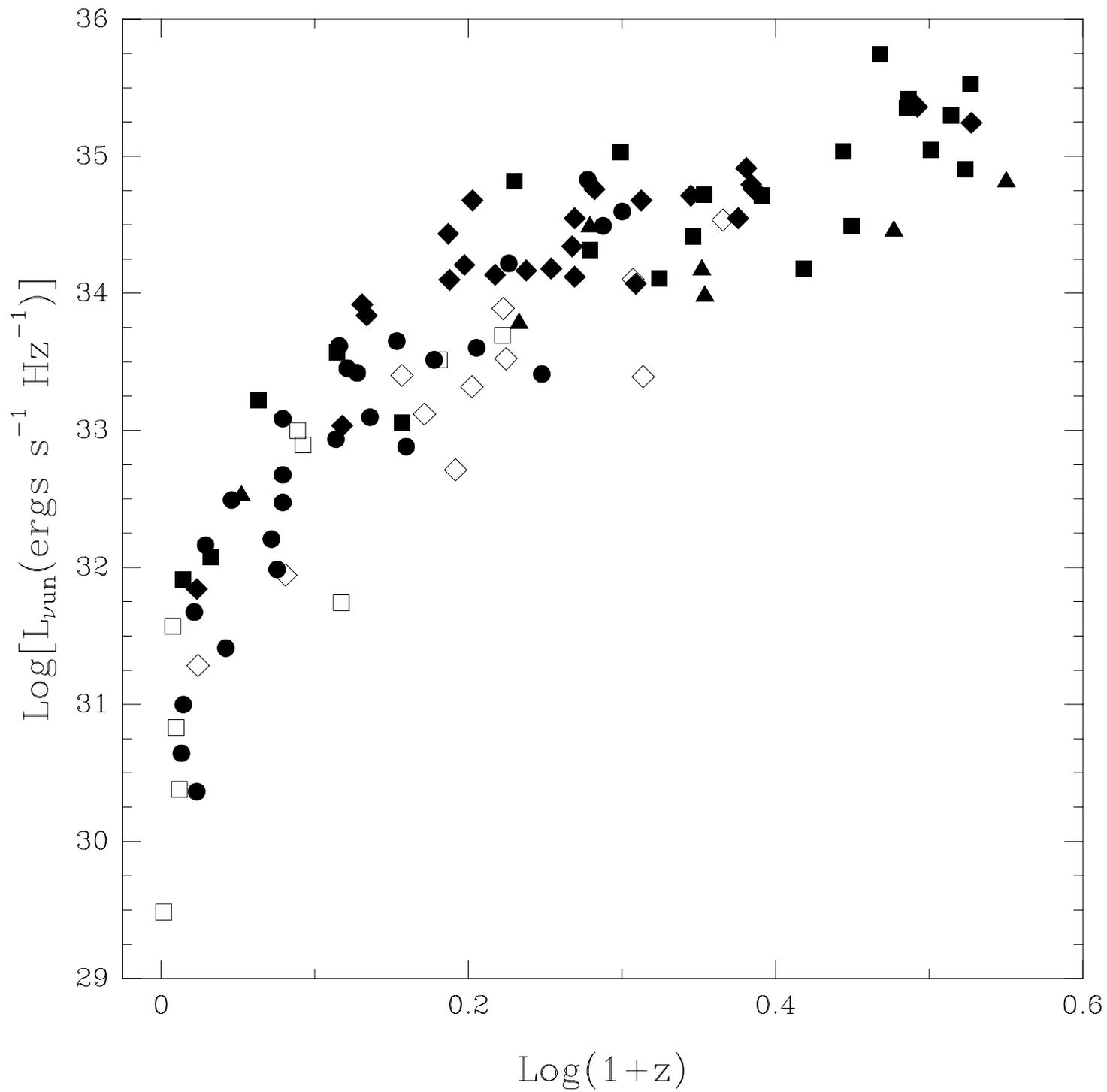
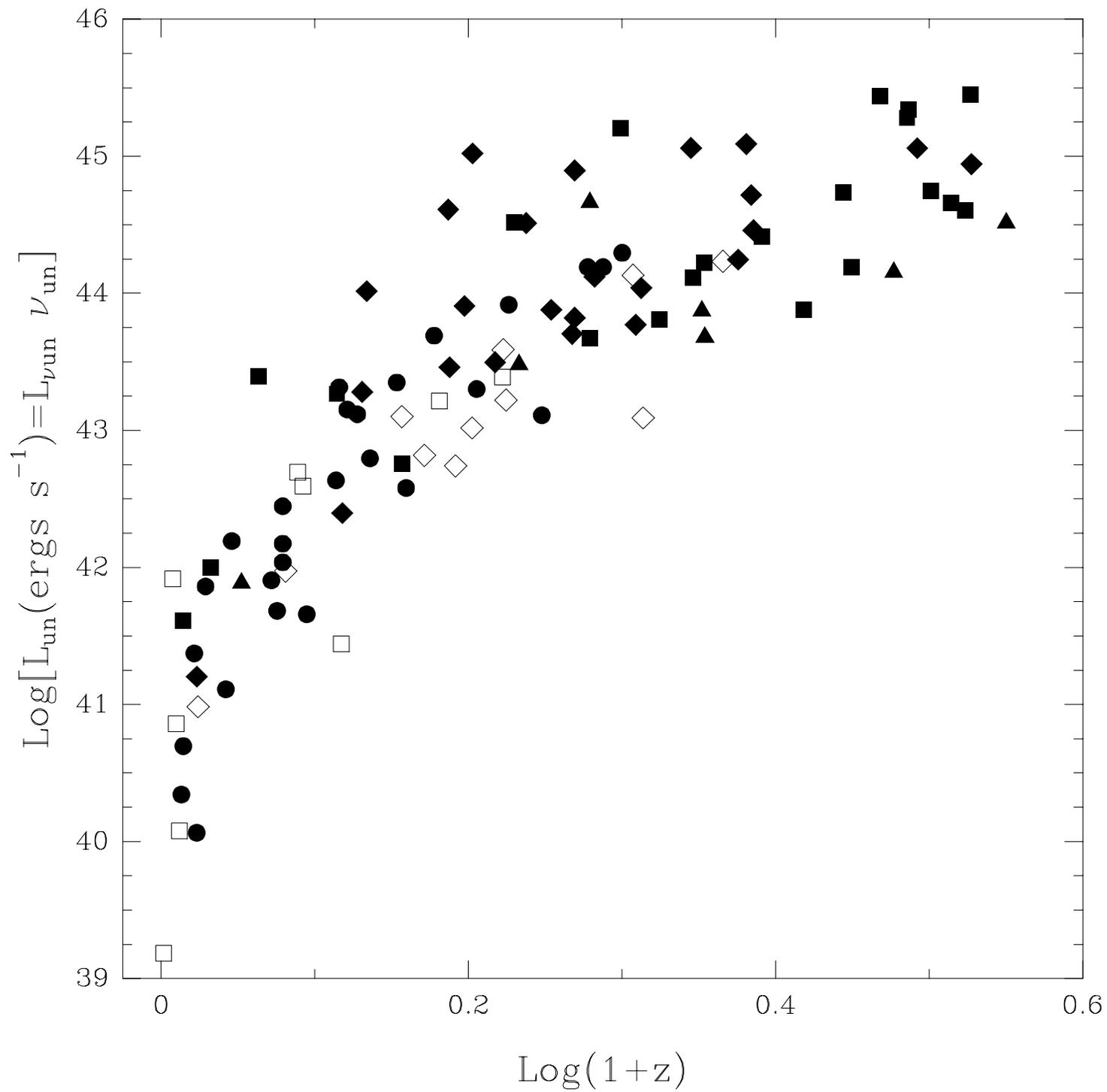
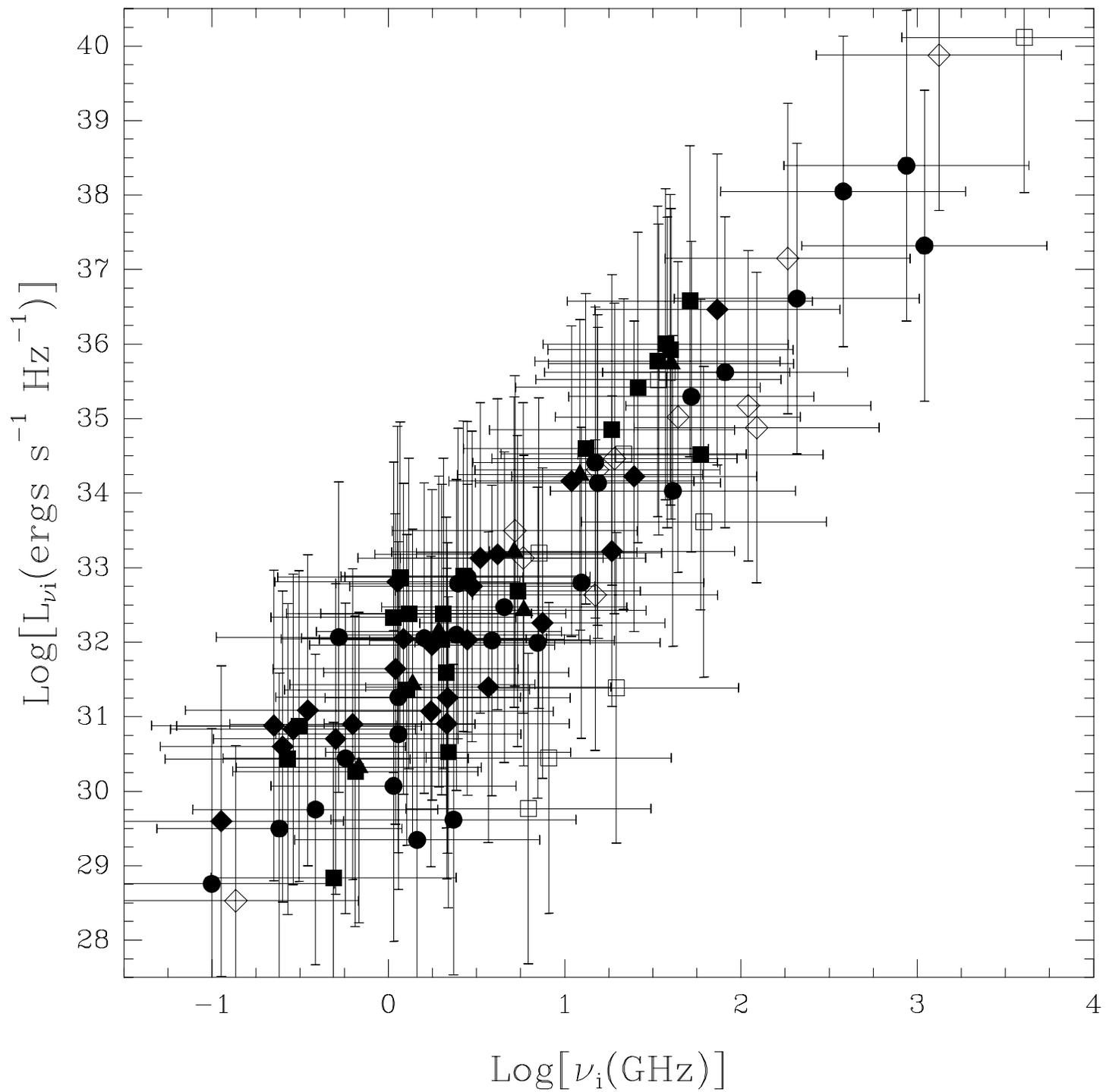
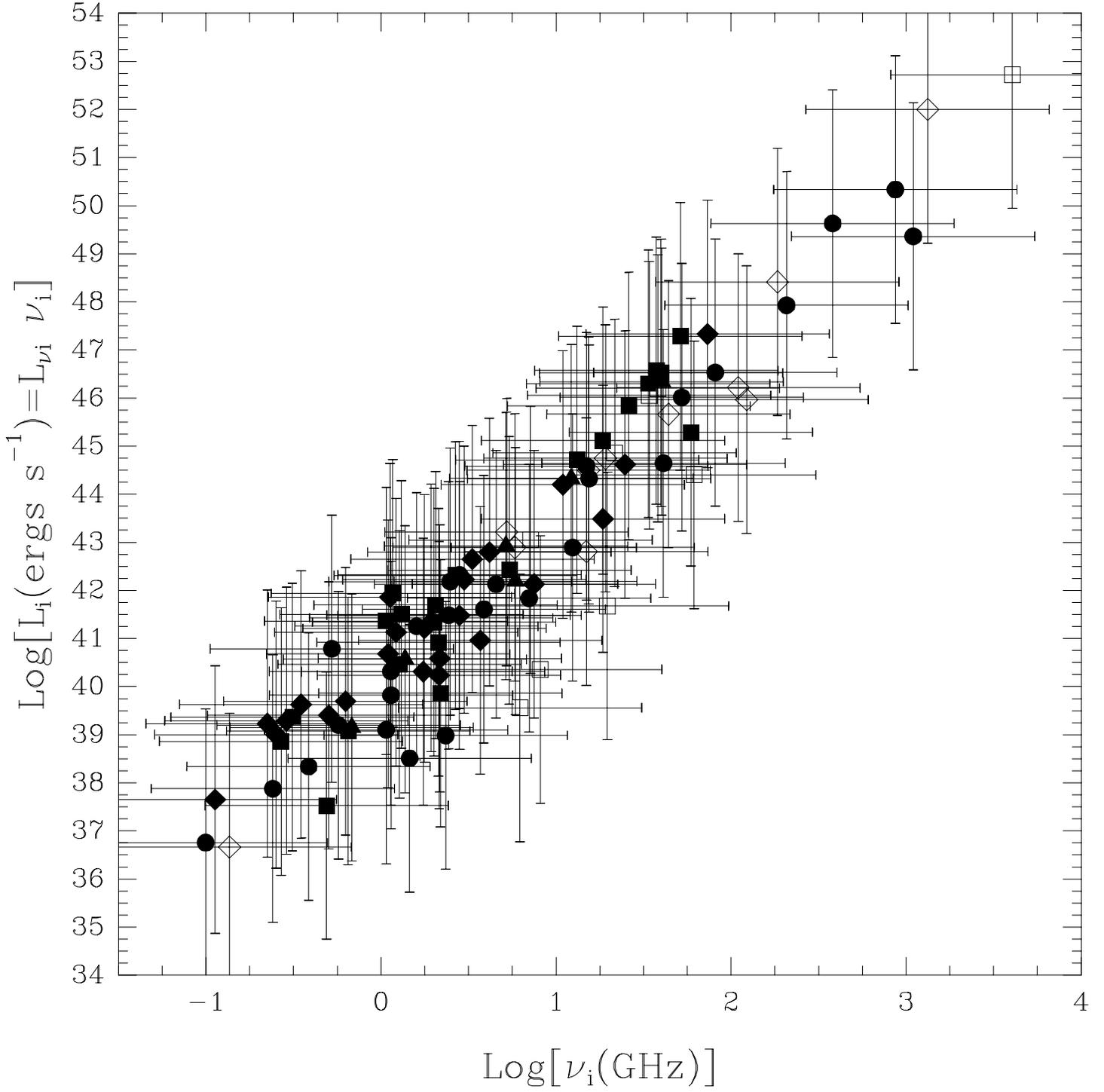
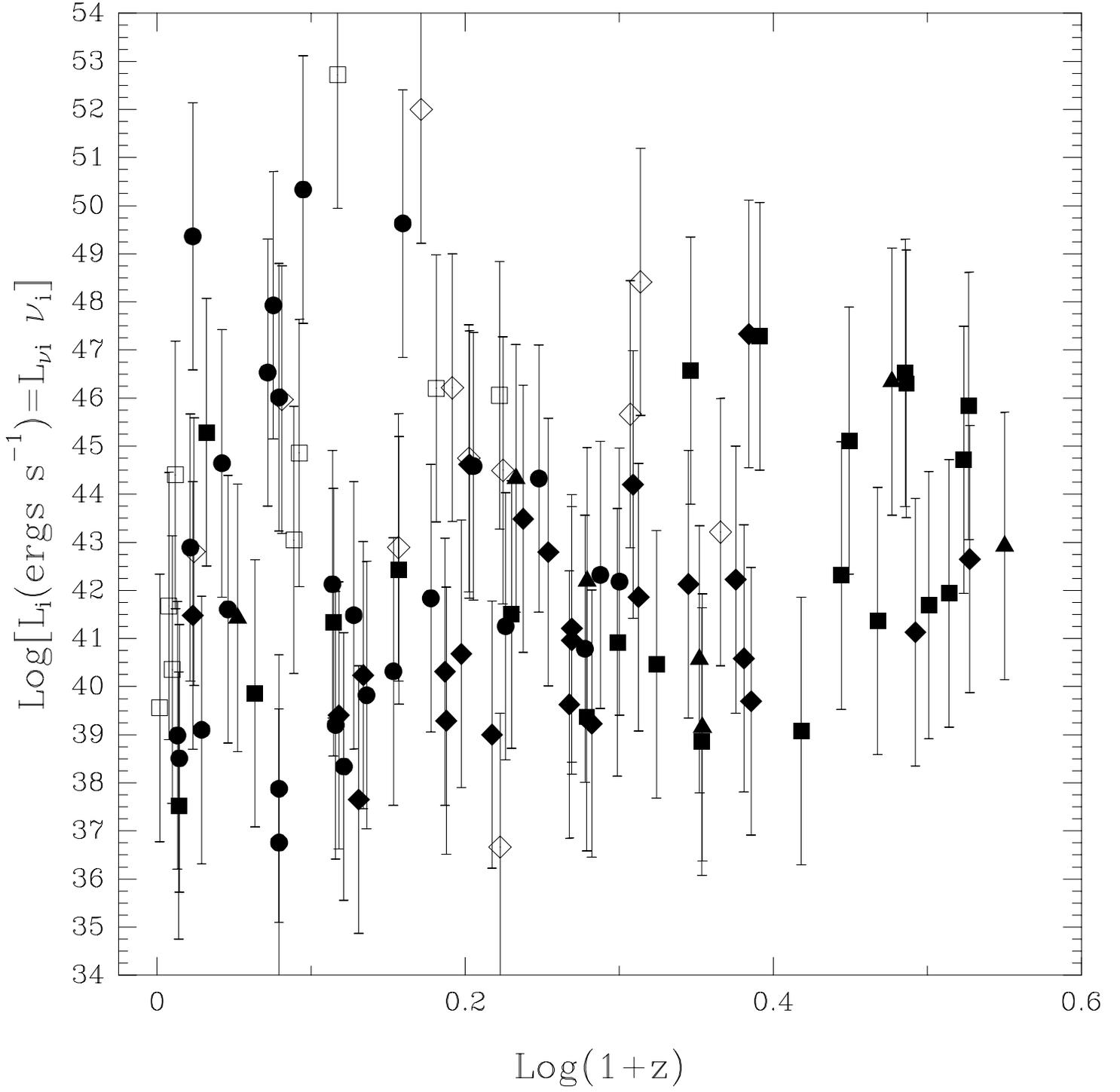
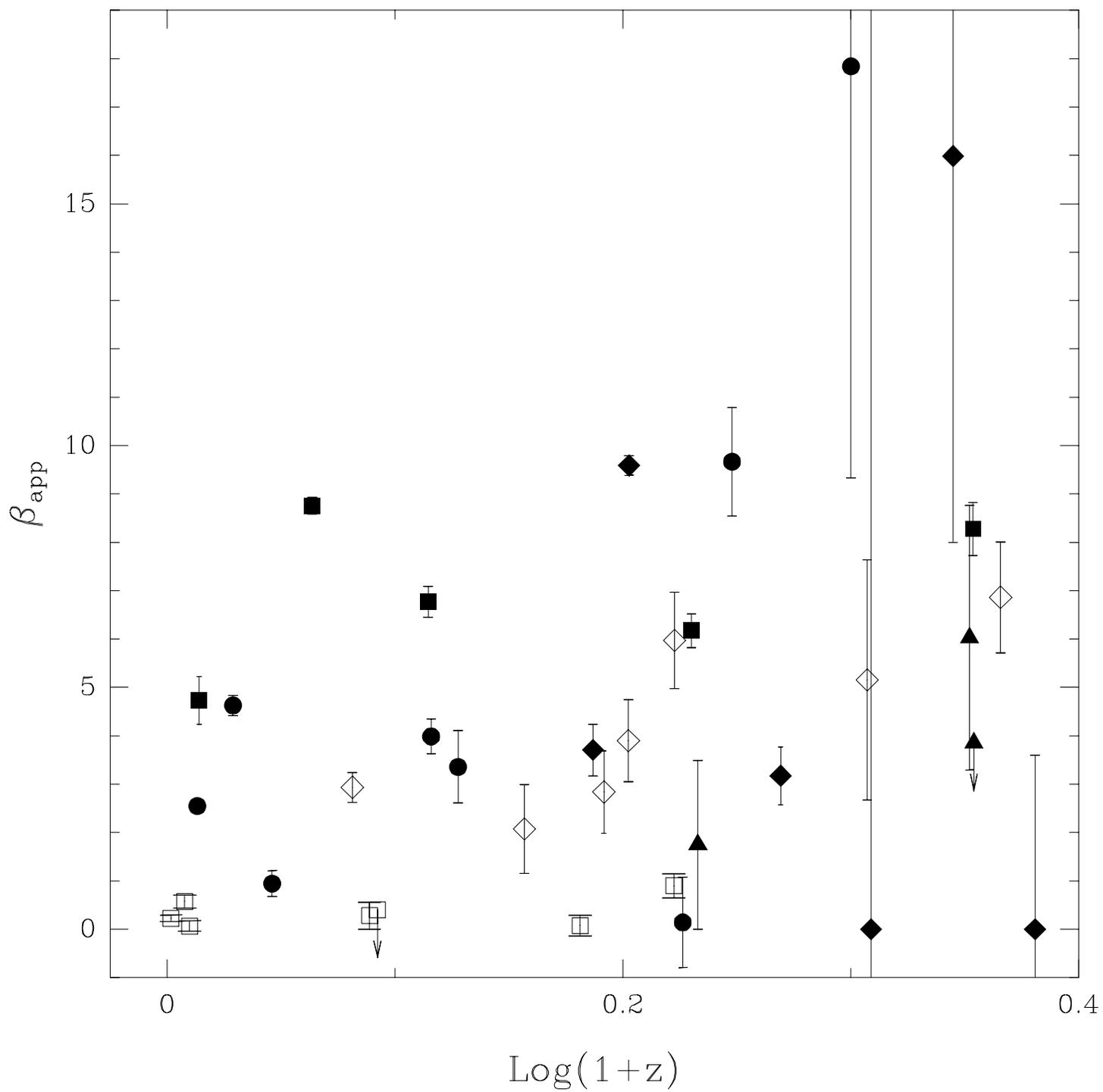
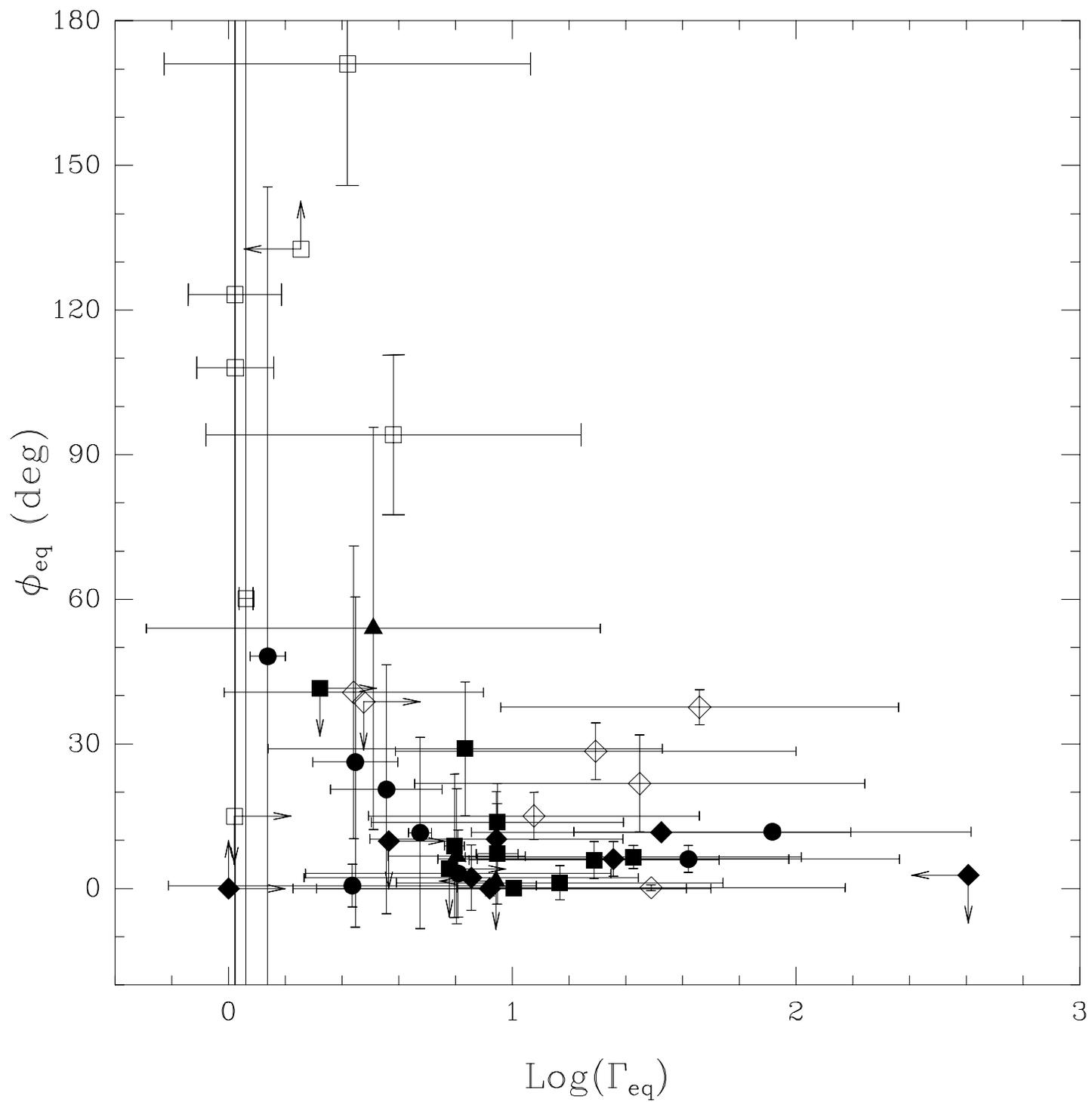
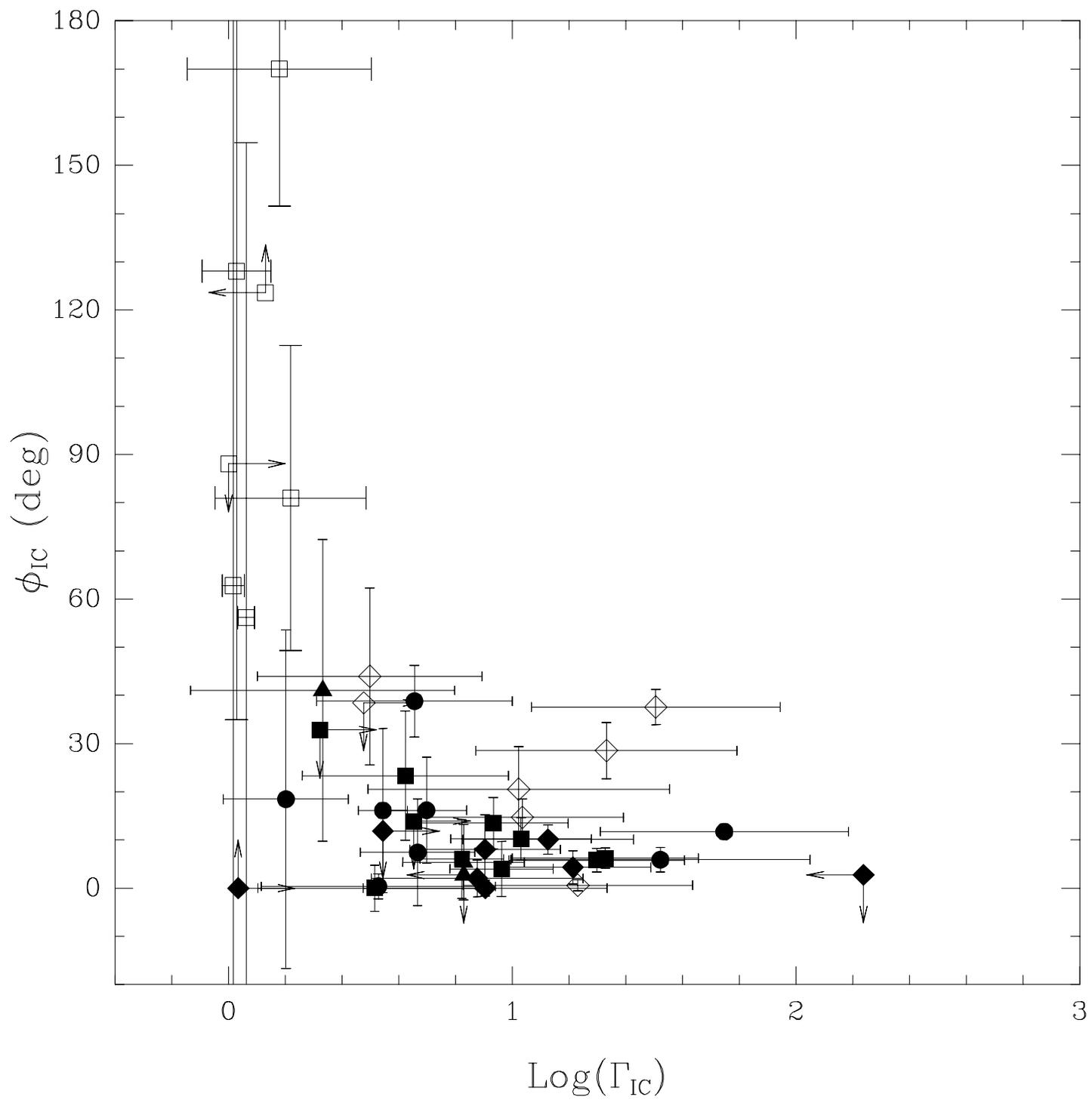
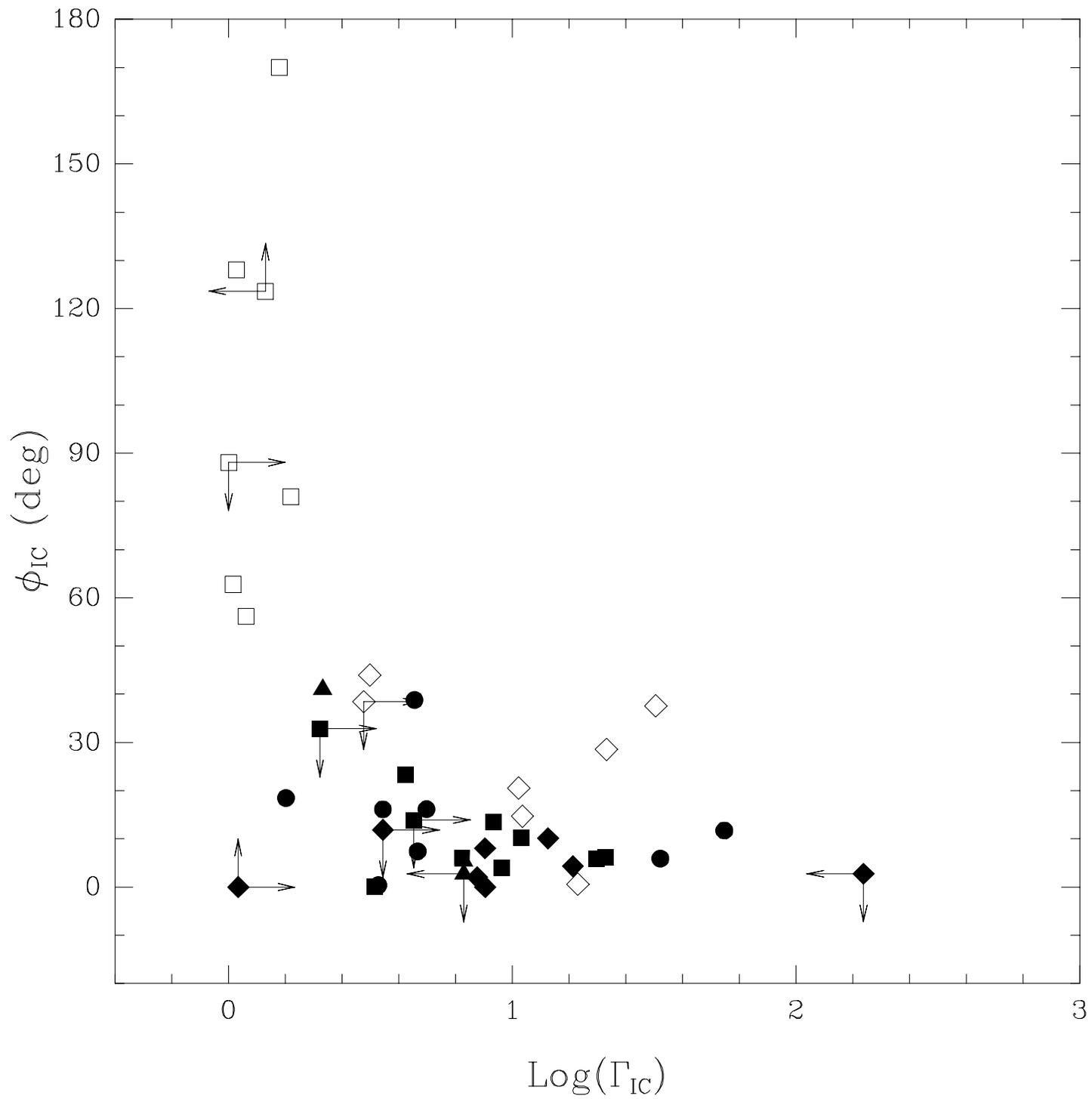
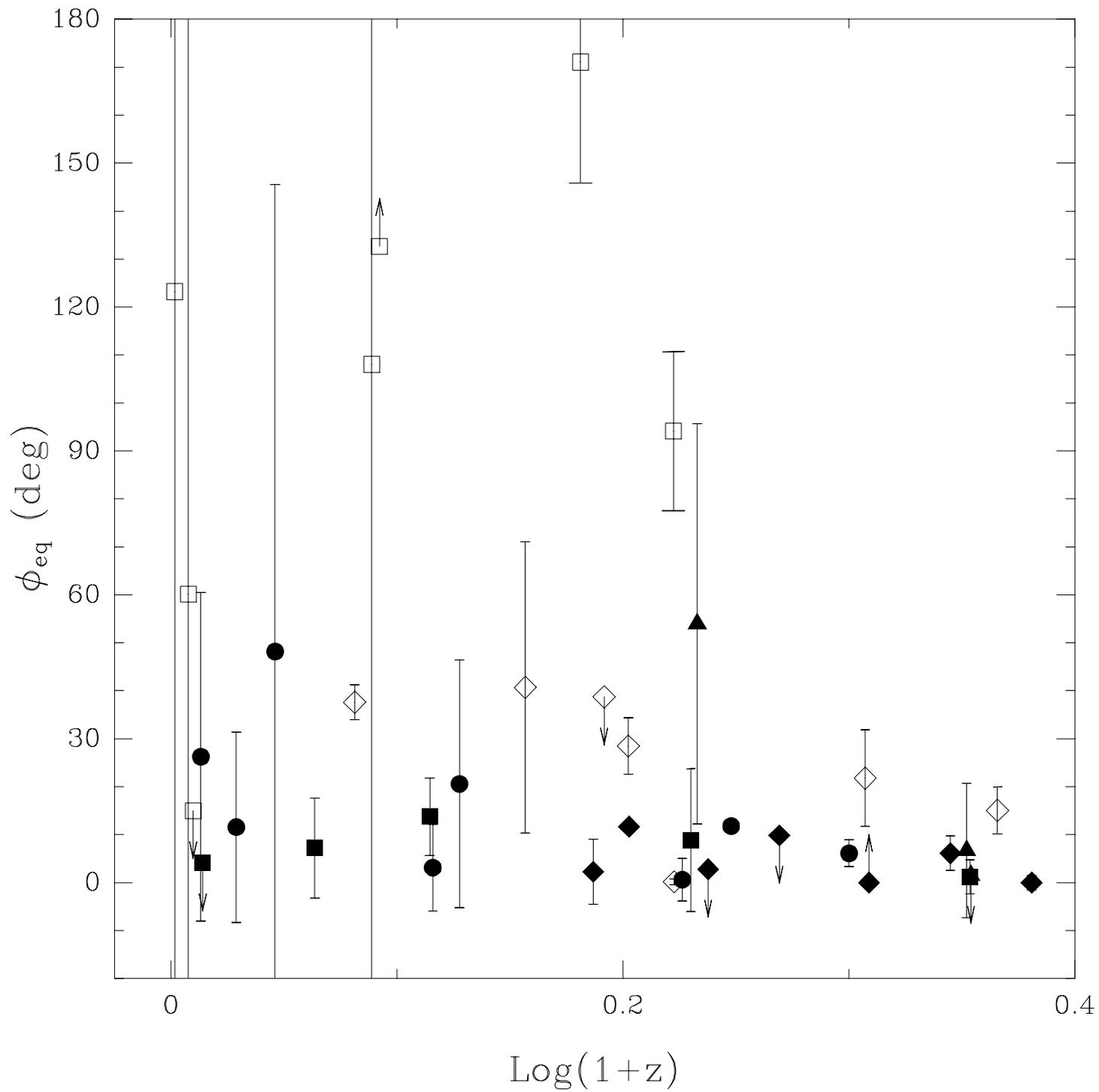
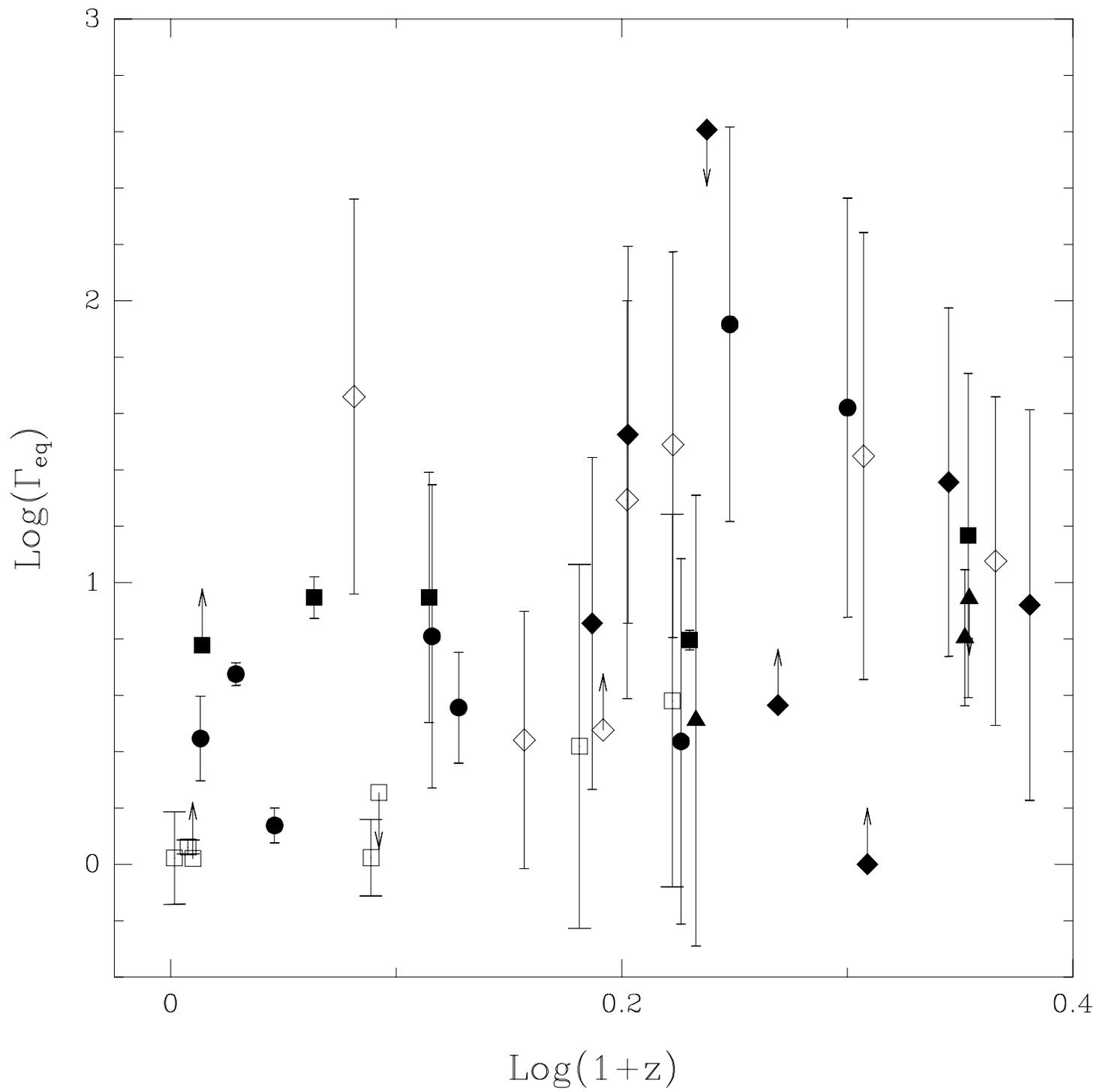
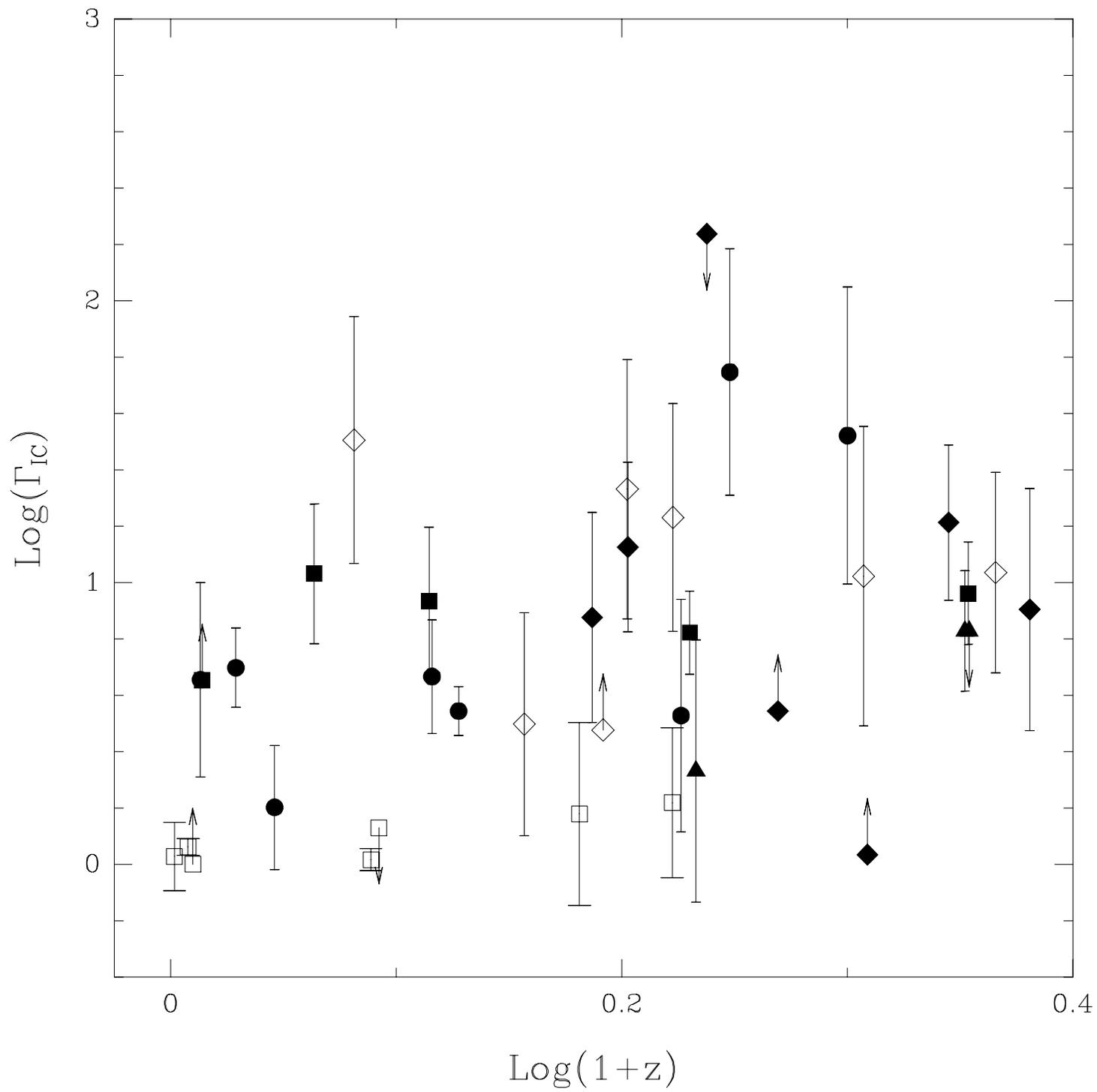
\end{document}